\documentclass[prb,aps,10pt,twocolumn,nofootinbib,superscriptaddress]{revtex4-1}
\usepackage{amsmath}
\usepackage{amssymb}
\usepackage{graphicx}
\usepackage{hyperref}
\usepackage{framed}
\usepackage{comment}
\usepackage{pifont}
\usepackage{epstopdf}
\usepackage{amsthm}
\usepackage{multirow}
\usepackage[caption=false]{subfig}
\usepackage{color,colortbl}
\newcommand{\cmark}{\ding{51}}
\newcommand{\xmark}{\ding{55}}
\newcommand{\eqnref}[1]{Eq.~(\ref{#1})}
\newcommand{\ket}[1]{|#1\rangle}
\newcommand{\bra}[1]{\langle#1|}

\newcommand{\cat}[1]{\mathbf{#1}}
\newcommand{\U}{\mathrm{U}}
\newtheorem{lemma}{Lemma}

\definecolor{Gray}{gray}{0.9}

\usepackage{tikz}
\usetikzlibrary{external}
\usetikzlibrary{decorations.markings}
\usetikzlibrary{arrows.meta}
\usetikzlibrary{calc}
\begin{document}

\title{Cheshire charge in (3+1)-D topological phases
}

\author{Dominic V. Else}
\affiliation{Physics Department, University of California,  Santa Barbara, California 93106, USA}

\author{Chetan Nayak}
\affiliation{Station Q, Microsoft Research, Santa Barbara, California 93106-6105, USA}
\affiliation{Physics Department, University of California,  Santa Barbara, California 93106, USA}

\begin{abstract}
We show that $(3+1)$-dimensional topological phases of matter generically support loop
excitations with topological degeneracy. The loops carry ``Cheshire charge":
topological charge that is not the integral of a locally-defined topological
    charge density. Cheshire charge has previously been discussed in non-Abelian
    gauge theories, but we show that it is a generic feature of all (3+1)-D
    topological phases (even those constructed from an Abelian gauge group).
Indeed, Cheshire charge is closely related to non-trivial three-loop braiding.
We use a dimensional reduction argument to compute the topological degeneracy
of loop excitations in the $(3+1)$-dimensional topological phases associated with
Dijkgraaf-Witten gauge theories. We explicitly construct membrane operators associated
with such excitations in soluble microscopic lattice models in
${\mathbb{Z}_2}\times{\mathbb{Z}_2}$ Dijkgraaf-Witten phases and generalize
this construction to arbitrary membrane-net models.
We explain why these loop excitations are the objects in the braided fusion 2-category
$Z(\cat{2Vect}_G^{\omega})$, thereby supporting the hypothesis that $2$-categories
are the correct mathematical framework for $(3+1)$-dimensional topological phases.
\end{abstract}
\maketitle

\section{Introduction}
\label{sec:intro}

Emergent particle-like topological excitations are the
hallmark of $(2+1)$-dimensional topological phases of matter (see Ref. \onlinecite{Nayak2008} and references therein),
which can be classified through the braiding statistics of these excitations. The set of particle types, and their fusion and braiding properties (as encapsulated by $F$- and $R$-matrices satisfying certain consistency conditions, such as the pentagon and hexagon identities) form an algebraic structure
called a unitary modular tensor category (see, e.g. Appendix E of Ref. \onlinecite{Kitaev2006a}).
This algebraic structure, together with the
thermal Hall conductance, actually is sufficient to determine \emph{all} the universal properties of a topological phase, not just those relating to the particle-like excitations.
An important example that has been the focus of recent interest is
the ground state degeneracy on surfaces with handles and boundaries\cite{Beigi11,Kapustin2011,Kitaev2012,Lindner2012,Clarke2013a,Wang2015a,Barkeshli13b,Barkeshli13c,Barkeshli13d,Levin2013,
Kapustin2014a,Lan2015,Yoshida15,Cong16}. Recent work
has exploited
the fact that even Abelian topological phases
\footnote{When a collection of quasiparticles at fixed positions has a unique state on a disk, braiding
operations result in the accrual of phases; such topological states are called Abelian.
When there is topological degeneracy associated with quasiparticles, braiding transformations
need not commute, and such phases are called non-Abelian.}
can have fully-gapped boundaries that admit topologically-degenerate ground states.
As in the case of the topological degeneracy associated with a collection of non-Abelian
particle excitations on a disk, this degeneracy cannot be broken by local perturbations
and, therefore, forms a decoherence-free subspace of the Hilbert space,
with potential applications to quantum information processing.

It is natural to expect that, in $(3+1)$-dimensions, topological phases will be classified by the
braiding properties of both particle-like and loop-like excitations.
These properties have been explored in Refs. \onlinecite{WangChe2014,Jian2014,Bi14,Jiang14,Lin15,Wang15,
WangChe15,Chen16,Putrov2016}.
It has been shown
that particle-loop braiding, loop-loop braiding, and three-loop braiding are important
invariants characterizing $(3+1)$-dimensional topological phases \cite{WangChe2014,Jian2014}.
However, it is not clear that these data are a complete characterization. In fact, as we will show,
even the more basic problem of determining the set of
loop-like excitations has not been completed for known $(3+1)$-dimensional topological phases
associated with Dijkgraaf-Witten gauge theories \cite{Dijkgraaf1990}.
In this paper, we show that this set generally includes loop-like excitations that carry topological
charge that is locally unobservable, often (somewhat fancifully) called Cheshire charge \cite{Alford1990,Preskill1990,Bucher1991}.
It was previously recognized that flux loops could carry charge \cite{WangChe2014,Jian2014},
but the different charge states of those flux loops excitations were not degenerate. Our Cheshire-charge-carrying
loop excitations are a topologically-distinct class of excitations whose different charge states must have the same energy.
In fact, we will argue that loops exhibiting non-trivial three-loop braiding \emph{necessarily} also carry Cheshire charge.
The associated topological degeneracy of loop-like excitations in $(3+1)$-dimensional topological phases
is a cousin of the topological degeneracy of gapped edges of $(2+1)$-dimensional topological phases, and can be mapped to the latter by dimensional reduction. However, it is worth emphasizing that
the presence of Cheshire charge gives a simple, fully $(3+1)$-dimensional explanation
of non-trivial three-loop braiding.

Loops carrying Cheshire charge are, in fact, the objects in a mathematical
structure that is, potentially, a framework for understanding and classifying
$(3+1)$-dimensional topological phases: braided fusion $2$-categories. The basic data
in such a description are the natural generalization of the $F$ and $R$ symbols of
$(2+1)$-dimensional topological phases. They determine, in principle, all the topological properties of a $(3+1)$-dimensional topological
phase -- including, but not limited to, three-loop braiding and topological degeneracies.
These data satisfy stringent consistency conditions, generalizing the pentagon and hexagon
identities. Indeed, our initial motivation for this work was that solving these equations for Dijkgraaf-Witten gauge theories turns out to require the introduction of extra structure, for which Cheshire charge is the physical interpretation. Thus, one might say (with tongue firmly in cheek): Cheshire charge is a \emph{falsifiable prediction} of
braided fusion $2$-categories that is confirmed by a more direct calculation.

In Section \ref{CC-in-DW}, we use the dimensional reduction procedure introduced in
Ref. \onlinecite{WangChe2014} and described briefly above to demonstrate the existence of
loop-like excitations carrying Cheshire charge. In Section \ref{dimreduction_enumerate},
we compute the topological degeneracy of
loop-like excitations in Dijkgraaf-Witten phases. In Section \ref{sec:lattice-model}, we construct
the operators that create loop excitations with Cheshire charge in a soluble lattice model
of ${\mathbb{Z}_2}\times{\mathbb{Z}_2}$ Dijkgraaf-Witten phases constructed in
Ref. \onlinecite{Lin15}. This is generalized to arbitrary membrane-net models in Section \ref{membranenet}.
In Section \ref{sec:2-categories}, we explain our results from the category theory perspective:
we have, in fact, been constructing the objects in the braided fusion 2-category
$Z(2\text{Vect}_G^\omega)$, where $Z$ denotes the center functor, and $2\text{Vect}_G^\omega$
is the fusion 2-category of $G$-graded 2-vector spaces, twisted by a 4-cocycle $\omega$.

\section{Cheshire charge as a ``non-Abelian'' fusion rule}
A helpful way to think about the Cheshire charge that we will discuss is in terms of fusion rules. We can make an analogy to the case of topological degeneracy of point excitations in (2+1)-D. Such excitations have a ``fusion rule'' $a \times b = \bigoplus_{c} N_{ab}^c c$ which tells us the possible particle types that result when treating two particles $a$ and $b$ as a single composite particle. If the ``fusion rule'' is non-Abelian, which is to say that there is more than one possible outcome on the right-hand side of the fusion rule, then when the particles are well-separated this manifests as a topological degeneracy. For example, consider the case of Ising anyons, with particle types $\{ 1, \sigma, \psi \}$ and fusion rules
\begin{equation}
\psi \times \psi = 1, \,\,\, \sigma \times \psi = \sigma, \,\,\, \sigma \times \sigma = 1 + \psi.
\end{equation}
Then if we have two $\sigma$ particles, their joint state could look like the vacuum or like a $\psi$ particle. Moreover, if the two particles are well-separated, then states corresponding to the $1$ and $\psi$ fusion channels are locally indistinguishable, and hence degenerate. (Strictly speaking, in a finite system without boundary, then the overall state of the system must fuse to the vacuum and so the $\psi$ channel is not allowed when there are only two particles. But by considering \emph{four} $\sigma$ particles, for example, one gets a two-fold degeneracy even with the constraint that the overall state must fuse to vacuum). On the other hand, if we move the particles so that they are close to each other then the fusion channel can be measured locally.

The topological degeneracy of Cheshire charge loops can also be understood in the same way. A closed loop excitation looks from far away like a point excitation. So one can define a ``fusion rule'' that asks what point excitations types one can thus obtain, from a given loop excitation type. For example, in the next section we will explicitly construct, in the case of the 3-D toric code, a loop excitation c such that this ``fusion rule'' takes the form.
\begin{equation}
c \to 1 + e, \label{cfusion}
\end{equation}
where $e$ is the (unique) non-trivial point particle in the 3-D toric code. Thus, even though the 3-D toric code can be considered as a gauge theory for an Abelian $\mathbb{Z}_2$ gauge theory, the fusion rule of excitations is non-Abelian.
Moreover, as in the 2-D case, we shall see that the states corresponding to the different ``fusion channels'' are locally indistinguishable and thus degenerate. (Again, to actually see the degeneracy in a finite system without boundary we would need to consider two loops, because a single loop is subject to the constraint that the total state must fuse to the vacuum, so the $e$ channel in \eqnref{cfusion} is not allowed.)

\section{Cheshire charge in the 3D toric code}
\label{cheshire_3d_toric}
Here we consider the 3D toric code, which is the simplest possible (3+1)-D topological phase, and show that it exhibits a Cheshire charge phenomenon. The 3D toric code consists of spin-1/2 particles living on the links of a 3D lattice, with Hamiltonian
\begin{equation}
\label{H3dtoric}
H = -\sum_i A_i - \sum_p B_p,
\end{equation}
where
\begin{equation}
A_i = \prod_{j \sim i} \sigma^x_{ij}
\end{equation}
\begin{equation}
B_p = \prod_{ij \in p} \sigma^z_{ij}
\end{equation}
and the first sum in \eqnref{H3dtoric} is over vertices of the lattice, and the second sum is over faces. A violation of the first term is called a ``charge'', and a violation of the second term is called a ``flux''.

If we choose a basis for the Hilbert space labelled by the eigenvalues $\sigma_i = \pm 1$ of the operators $\sigma^x_i$, then we can imagine that each such basis state represents a ``string-net'' configuration in which $\sigma_i = +1$ denotes the absence of a string, and $\sigma_i = -1$ the presence of a string, on the given link. The ground state of \eqnref{H3dtoric} is then the equal-amplitude superposition over all \emph{closed} string-net configurations, that is those configurations in which
in each vertex has an even number of strings going in.

Now let $l = i_1 i_2 \cdots i_{N+1}$ be some closed path such that $i_{k+1}$ is adjacent to $i_k$ for all $k$ and $i_{N+1} = i_1$. We now imagine turning off the first term of the Hamiltonian for all vertices $i_1, \cdots, i_N$, and instead turning on the term
\begin{equation}
\label{eqn:condense-e-particles}
-\sum_{k=1}^{N} \sigma^z_{i_k i_{k+1}},
\end{equation}
which creates a pair of point-like ``charge excitations'' of the original Hamiltonian at either
end of each link ${i_k i_{k+1}}$ belonging to path $l$. Thus, in eigenstates of the Hamiltonian,
each link resonates between a configuration in which these ``charge excitations'' are present
and one in which they aren't. In ground states, each link on path $l$ is in an equal-amplitude
superposition of the zero- and two-``charge excitation'' configurations. Thus, there is
equal amplitude for any vertex in $l$ to be the endpoint of a string or to not be the endpoint of
a string, i.e. the charges condense on $l$.
It is easily seen that the new Hamiltonian $H^{\prime}$ has a two-fold degnerate ground state. In fact, each term of the Hamiltonian is minimized by both $\ket{+}$ and $\ket{-}$, where $\ket{+}$ ($\ket{-}$) is the superposition over all string-net configurations such that strings terminate only on the path $l$, and such that there are an even (odd) number of terminations on $l$. In the latter case, the other end of at least one of the strings must terminate on
a charge excitation elsewhere in the system. While this would cost energy if the other end were an electric excitation
of the bulk, there would be no additional energy cost if the other end were on a second loop of the form
(\ref{eqn:condense-e-particles}), i.e. the energy would be the same regardless of whether an even or odd number of
strings went from one loop to the other. In the two resulting states, both loops
carry the same overall charge, which is $+1$ or $-1$. The charge on a loop cannot be distinguished or connected by operators that do not act over the entire loop. Thus, the degneracy is ``non-local'', analogous to the ground-state degeneracy in the 2D (or 3D) toric code. This is a simple manifestation of the phenomenon known as ``Cheshire charge''.
Cheshire charge has previously been studied in the context
of non-Abelian  gauge theories \cite{Alford1990,Preskill1990,Bucher1991}.
Here, we have shown that it is also present in $3+1$-D gauge theories with finite Abelian gauge group.

In this example one can see two of the properties that we expect to be true of Cheshire charge in general:

\emph{Charge condensation}.-- One way to think of the loop excitation carrying Cheshire charge is that charge has \emph{condensed}, but only on the loop $l$. A consequence of this is that a charge can be absorbed by a loop, leaving behind no local excitation apart from the loop itself.

\emph{Cheshire-flux opacity}.-- In the 3D toric code one can construct a \emph{membrane operator} $\Sigma_S = \prod_{ij \perp S} \sigma^x_{ij}$, where $S$ is a surface in the dual lattice, and the product is over all links $ij$ intersecting $S$. This membrane operator commutes with the original Hamiltonian $H$ (except at the boundary of $S$), and its effect is to move flux excitations around. On the other hand, if the surface $S$ intersects the loop $l$ an odd number of times, then $\Sigma_S$ does \emph{not} commute with the new Hamiltonian $H^{\prime}$. This has a physical consequence: a flux loop cannot be moved through a Cheshire charge loop without leaving behind a disturbance; Cheshire charge loops are \emph{opaque} to flux loops.

\begin{figure*}
\includegraphics[width=14cm]{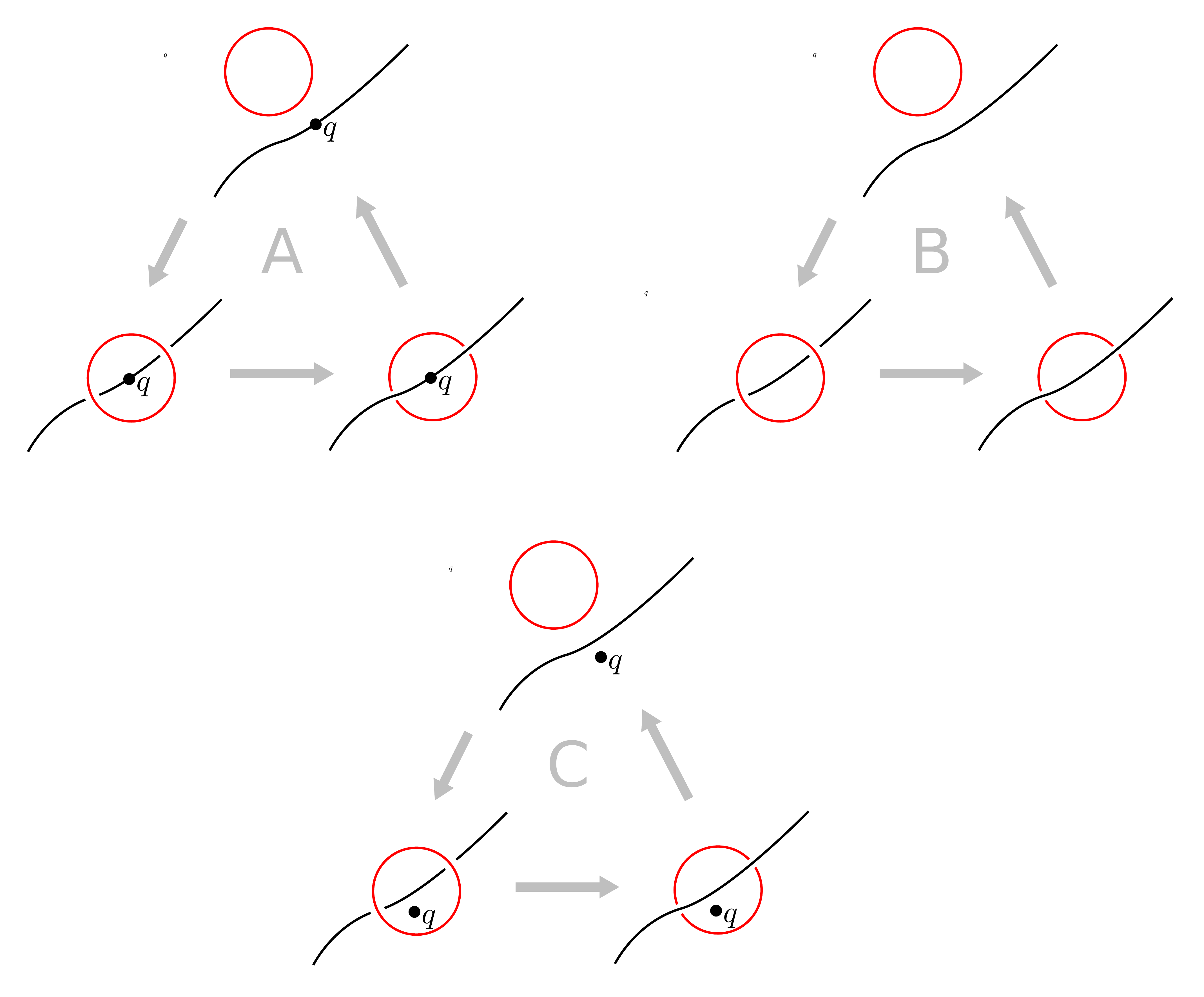}
\caption{\label{opacity_consistency}Processes A, B and C must have the same amplitude. Here the red loop is a flux loop and the black loop is a Cheshire charge loop.}
\end{figure*}

One might wonder whether this problem could be avoided simply by redefining the membrane operator $S$ near $l$. However, it turns out that this is impossible due to a topological obstruction. Indeed, suppose it were possible. Then we could imagine the process depicted in Figure \ref{opacity_consistency}, in which a flux loop is braided around a point charge sitting on a Cheshire charge loop (this process is only allowed if flux loops can pass through Cheshire charge loops); call this process A. Since the point charge can be annihilated by a local operation, process A must give the same amplitude as another process (B) in which the point charge is not present. Similarly, the same amplitude must result for a process (C) where the point charge is located slightly away from the Cheshire charge loop. But process C is topologically equivalent to process B followed by braiding the point charge around the flux loop. As this gives an extra phase factor of -1, we reach a contradiction.

The term (\ref{eqn:condense-e-particles}) induces Cheshire charge along the loop $l$ in the ground state.
We could modify this term to induce Cheshire charge along an arc or a set of arcs; in the absence of flux through the loop, there
is no reason for the loop $l$ to be closed. Indeed, the Cheshire charge need not even be confined to a one-dimensional
subsystem. The term (\ref{eqn:condense-e-particles}) could, instead, be turned on over some surface or even a
three-dimensional blob. In this paper, we will focus on loop-like excitations. As we shall see, flux loops that support Cheshire
charge will play a special role in the topological description of charge and flux excitations of 3D topological phases.
It may be the case that more general objects, such as surfaces, blobs, etc. can be understood as conglomerations
of point-like and loop-like excitations, just as finite-energy loops and 2D blobs can, presumably, be understood
as conglomerations of point-like excitations in 2D topological phases.

\section{Cheshire charge in more general models and Three-Loop Braiding}
\label{sec:cheshire-charge-general}

In this section, we will argue that Cheshire charge is intimately related
to non-trivial three-loop braiding
\cite{WangChe2014,Jian2014,Bi14,Jiang14,Lin15,Wang15,WangChe15,Chen16}.
First, we generalize the above considerations to formulate a theory of Cheshire charge in a more general (3+1)-D topological phase. For simplicity we will assume that particle-particle fusion and particle-loop braiding is Abelian. We will let $A$ be the set of particle types. Then for any particle $a \in A$ and loop excitation $l$, there is some phase factor $\chi_l(a)$ associated with braiding a particle $a \in A$ around the loop $l$. For consistency with particle fusion, we require that $\chi_l$ be linear in $A$, i.e.\ $\chi_l(a_1) \chi_l(a_2) = \chi_l(a_1 a_2)$. Now, by analogy with the toric code example considered above, we will assume that the allowed values of Cheshire charge carried by $l$ correspond to the subgroup $A_l \leq A$ of particles which can be locally annihilated on $l$. We will call this the \emph{condensation group} of the loop excitation $l$.

By the topological arguments discussed for the toric code above, we can then draw a relationship between the ability of loop excitations to pass through each other and the Cheshire charge which they carry. Specifically, if two loop excitations $l$ and $l'$ can pass through each other, we must have the compatibility condition
\begin{framed}
$\chi_l(a) = 1$ for all $a \in A_{l'}$ and $\chi_{l'}(a) = 1$ for all $a \in A_{l}$.
\end{framed}
One can then conjecture that this is the \emph{only} required condition; that is, if the condition
in the box above is satisfied so that there is no topological obstruction from charge condensation, then the two loops can pass through each other. This conjecture will be true for all theories we consider in this paper.

There is a very important consequence to this conjecture. In the toric code example above, Cheshire charge loops are distinct from the flux loop excitations. However, in a general topological phase this will not be the case. In fact, in a general phase there can be a non-trivial topological amplitude associated with a ``three-loop'' braiding process in which two loops $l_1$ and $l_2$ are braided around each other while they are both linked with a third loop $l$. This implies that it must not be possible to pass both $l_1$ and $l_2$ through $l$ without disturbance (because otherwise the base loop $l$ could be unlinked from the others without changing the topological amplitude). If the conjecture is true, the following holds:
\begin{framed}
There is a non-trivial amplitude for three-loop braiding of $l_1$, $l_2$, and $l$
iff at least one of $l$,$l_1$, or $l_2$ carries Cheshire charge.
\end{framed}
It is tempting to speculate that fractional Cheshire charge can also occur.
The argument described in Sec. \ref{CC-in-DW} finds fractional charges
attached to the dimensionally-reduced versions of flux loops, which are then dyonic
particles \cite{WangChe2014}, as described in Appendix \ref{sec:degeneracy-from-topo}.
If this speculation is correct, then this fractional ``offset'' Cheshire charge would, in a
3D system that is large in every direction, spread out over the entire loop and would be
locally unobservable for the same reason as integer Cheshire charge. 
According to this line of reasoning,
when $l_1$ is linked with $l$, it is forced to carry a fractional Cheshire charge, in addition to
any integer Cheshire charge that it may carry according to the considerations in this paper.
The non-trivial amplitude for braiding $l_1$ with $l_2$ that is also linked with
$l$ would then be nothing more than the charge-flux Aharonov-Bohm effect
for the associated fractional Cheshire charges. Though it is suggestive, justifying this
picture is beyond the scope of this paper.

We close by noting that a loop excitation need not have the same condensation group everywhere on the loop. Instead,
a loop might be subdivided into arcs, each of which has a different condensation group. Then the boxed condition above
must hold for at least one condensation group in each loop, $A_l$ and $A_{l'}$, in order for the loops to
pass through each other. In the case of the toric code, we could have condensed charge-like excitations
in Eq. (\ref{eqn:condense-e-particles}) along only half of a loop, rather than the whole loop. More interestingly,
we could condense charge-like excitations along two disjoint arcs, in which case even a single loop would have non-trivial
degeneracy since both arcs could even or odd charge while the whole system would have vanishing charge. (Each arc must be long, or else there will be an energy splitting between these two states.) Similar considerations
hold for our more general classification below: there will be loop excitations in which the loop is subdivided into sections,
each of which is characterized by different topological data, subject to consistency conditions.

\section{Cheshire charge in Dijkgraaf-Witten theories}
\label{CC-in-DW}

In this section, we specialize to Dijkgraaf-Witten topological phases in (3+1)-D. A simple way to arrive at the excitation types is to consider a dimensional reduction argument, where we compactify one of the spatial dimensions, treating the resulting system as (2+1)-D. We must then address the question of what (2+1)-D phase results if we start from a (3+1)-D Dijkgraaf-Witten phase characterized by the 4-cocycle $\omega(g_1, g_2, g_3, g_4)$. The answer depends on which of the degenerate ground states of the (3+1)-D theory we started with, specifically on the gauge flux enclosed by the compactified dimension; for an Abelian  gauge group $G$ this flux is characterized by a group element $g \in G$. In Ref.~\onlinecite{WangChe2014},
it was found that the result of the dimensional reduction is the (2+1)-D Dijkgraaf-Witten phase corresponding to the 3-cocycle
\begin{multline}
\label{omega_x}
\widetilde{\omega}_g(g_1, g_2, g_3) = \omega(g, g_1, g_2, g_3)^{-1} \omega(g_1, g, g_2, g_3)^{-1}\\ \times \omega(g_1, g_2, g, g_3)^{-1} \omega(g_1, g_2, g_3, g).
\end{multline}
(this equation is sometimes referred to as the \emph{slant product}).

\begin{figure*}
\subfloat[Side view]{
\input{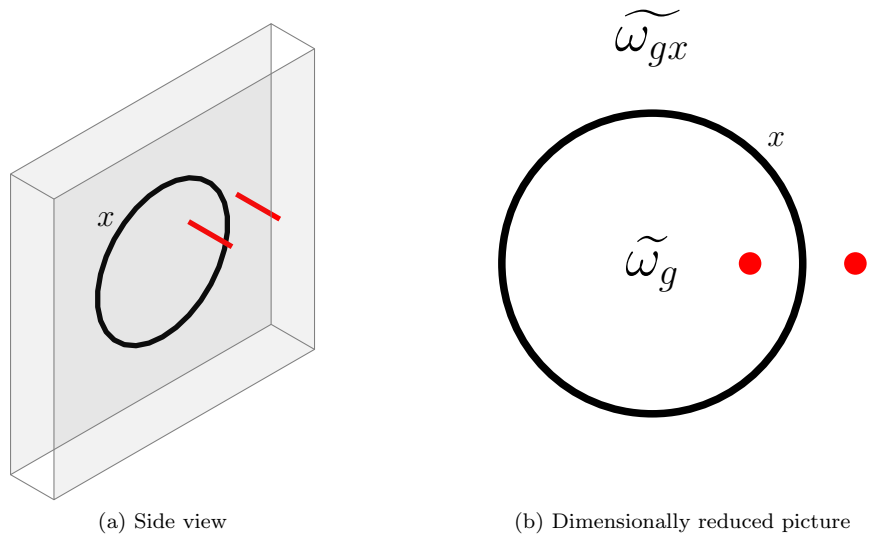}
}\hspace{2cm}
\subfloat[Dimensionally reduced picture]{
\raisebox{1.1cm}{
\begin{tikzpicture}
\draw[line width=0.1cm] (0,0) circle [radius=2];
\fill[red] (1.3,0) circle[radius=0.15];
\fill[red] (2.7,0) circle[radius=0.15];
\draw (0,0) node {\huge $\widetilde{\omega_g}$};
\draw (0,3) node {\huge $\widetilde{\omega_{gx}}$}; 
\draw (1.65,1.65) node {\large $x$};
\end{tikzpicture}
}
}
\caption{\label{dimred}Dimensional reduction of a loop excitation carrying flux $x$ in (3+1)-D. The small dimension in (a) is compactified (opposite surfaces are identified). The flux threaded through this compact dimension can be detected by moving a particle along a cycle (red lines). The two red lines in (a) differ by a braid around the $x$ flux, so they measure a flux differing by $x$.}
\end{figure*}
The above result pertains to the ground state of the topological phase. Now consider instead a loop-like excitation above the (3+1)-D ground state, carrying flux $x$. After dimensional reduction, this becomes a \emph{phase boundary} between two different (2+1)-D Dijkgraaf-Witten phases, with 3-cocycles $\widetilde{\omega}_{g}$ and $\widetilde{\omega}_{xg}$ respectively (see Figure \ref{dimred}). The properties of the excitation cannot depend on the choice of flux $g$, so without loss of generality we can set $g = 1$. One can show that $\widetilde{\omega}_1$ is necessarily a trivial 3-cocycle. Thus, we see that the loop-like excitation reduces to a phase boundary between the Dijkgraaf-Witten theory with 3-cocycle $\widetilde{\omega}_x$, and an ordinary (untwisted) gauge theory.


We can, therefore, classify the topological loop-like excitations
in a (3+1)-D Dijkgraaf-Witten phase (due to Cheshire charge)
from the degeneracy of gapped boundaries between
(2+1)-D Dijkgraaf-Witten phases.
When the dimensionally-reduced theory is also Abelian, we can use the classification of gapped boundaries of Abelian topological phases in 2+1-dimensions \cite{Beigi11,Kapustin2011,Kitaev2012,Lindner2012,Clarke2013a,Barkeshli13b,Barkeshli13c,Barkeshli13d,Levin2013,
Kapustin2014a,Yoshida15,Cong16}
to determine this degeneracy (see Appendix \ref{sec:degeneracy-from-topo}). 
In general, however, the dimensionally reduced theory might be non-Abelian, in which case understanding gapped boundaries becomes more challenging. Here we will take an alternative approach
by exploiting the relationship between Dijkgraaf-Witten topological phases and symmetry-protected topological (SPT) phases. The former are obtained by gauging the latter. The resulting Dijkgraaf-Witten theory is described by the same 4-cocycle $\omega(g_1, g_2, g_3, g_4)$ as the original SPT phase.

Thus, we see that the original dimensionally-reduced picture, in which the loop-like excitation becomes a boundary between the ground states of two different Dijkgraaf-Witten Hamiltonians, is equivalent to a new picture in which we have a boundary between the (2+1)-D SPT phase characterized by 3-cocycle $\widetilde{\omega}_x$ and the trivial SPT phase. Therefore, the loop excitations are classified by boundaries of SPT phases. In Appendix \ref{appendix_spt}, we give a more rigorous argument for the same conclusion, working directly in (3+1)-D so that we don't risk losing information in the dimensional reduction, and being careful as to the precise correspondence between the SPT phase and its gauged version. This approach also confirms that the degeneracy associated with the Cheshire charge loop does not depend on the global topology of space.

This has an important physical consequence in the case that $\widetilde{\omega}_x$ is non-trivial. The boundary of a non-trivial 2+1-D SPT phase always either is gapless or breaks the symmetry spontaneously (see, for instance, Ref. \onlinecite{Chen12}). We expect that in the simplest case (but perhaps not always) the original loop excitation in the (3+1)-D topological phase is gapped. Here, by ``gapped'', we mean that, in the presence of terms in the Hamiltonian that require
a loop excitation to lie along some specified curve, the Hamiltonian has an energy gap above its (possibly degenerate) ground state(s). Hence we arrive at the conclusion that the corresponding SPT boundary \emph{must break the symmetry spontaneously.} (In the original gauge theory this corresponds to the statement that near the loop excitation the gauge group is effectively Higgsed to a smaller group.)

At first glance, one might not think that this fact leads to any degeneracy in the original gauge theory, since a gauge theory that is Higgsed in the bulk does not carry degeneracy (except on a topologically non-trivial space such as a torus.) Here, the Higgsing takes place only along a loop, but it is still true that there is no degeneracy in the case of a single loop. Indeed, suppose that in the equivalent SPT system, a $\mathbb{Z}_2$ symmetry is spontaneously broken on the boundary, and there are two symmetry-breaking states $\ket{\uparrow}$ and $\ket{\downarrow}$ which get transformed into each other under this $\mathbb{Z}_2$. Then only their linear combination $\ket{+} = \frac{1}{\sqrt{2}}(\ket{\uparrow} + \ket{\downarrow})$ is $\mathbb{Z}_2$ invariant and corresponds to a state in the original gauge theory.

However, on further reflection this simply correponds to the fact that for a gauge thory on a closed manifold the total charge must be trivial.
Suppose that we instead have \emph{two} identical non-intersecting, unlinked loops $a$ and $b$ which both spontaneously break the symmetry in the same way. Then there are \emph{two} possible $\mathbb{Z}_2$ invariant linear combinations: $\ket{+}_a \ket{+}_b$ and $\ket{-}_a \ket{-}_b$, where $\ket{\pm} = \frac{1}{\sqrt{2}}(\ket{\uparrow} \pm \ket{\downarrow})$. Furthermore, we see that, provided that the loops are well-separated spatially, and sufficiently large (so that the symmetry-breaking states become exactly degenerate), these two states will be degenerate in energy. In fact, they are locally indistinguishable by any operators which commute with the global $\mathbb{Z}_2$. Since these are the only operators that correspond to physical operators in the original gauge theory, we conclude that there are \emph{two degenerate locally indistinguishable states} corresponding to a pair of unlinked loop excitations in the original (3+1)-D gauge theory. From this point forward in the paper, we will be slightly loose in our language and refer to a 2-fold (in the toric code case) topological degeneracy associated with each loop individually, remembering that we are supposed to project onto the subspace of zero total charge at the end.
This degeneracy corresponds to a locally unobservable Cheshire charge.

Finally, we emphasize that, in general, there will be more than one kind of excitation (potentially with different allowed values of Cheshire charge) carrying the same flux. In particular, since
any SPT phase (trivial or not) can potentially have spontaneous symmetry breaking at its edge, one can also construct loop-like excitations exhibiting Cheshire charge even if the dimensionally reduced cocycle $\widetilde{\omega}_x$ is trivial, for example if the original (3+1)-D gauge theory is untwisted ($\omega =1$) or if we are considering loops which carry no flux ($x = 1$); we saw an example of this for the 3D toric code in Section \ref{cheshire_3d_toric} above. On the other hand, in the case that $\widetilde{\omega}_x$ is trivial there will always be \emph{some} excitation carrying that flux without any Cheshire charge. We will deal with these issues in more detail in the next section.

\section{Enumerating the types of loop-like excitations}
\label{dimreduction_enumerate}

\subsection{General Result}

Now let us hone the above general considerations into a more precise description of all the possible loop-like excitations in an Abelian (3+1)-D Dijkgraaf-Witten theory with gauge group $G$ and 4-cocycle $\omega$.  As before, we identify a loop-like excitation carrying flux $x \in G$ with a boundary between the trivial SPT phase and the SPT phase characterized by 3-cocycle $\widetilde{\omega}_x$. Thus, the problem is reduced to enumerating all possible such boundaries.

First of all, we assume that $G$ is spontaneously broken (or, in the original gauge theory, Higgsed) down to a smaller group $H$ on the loop excitation. An SPT phase with respect to $G$ is still, after being spontaneously broken to $H$, an SPT phase with respect to $H$. However, in order for the edge to be gapped while $H$ is unbroken,
it must be a trivial SPT with respect to symmetry group $H$.
In other words, $(\widetilde{\omega}_x)_H$ must be a trivial 3-cocycle of $H$, where $(\widetilde{\omega}_x)_H$ denotes the restriction to $H$. Secondly, each of the symmetry breaking states could itself be a \emph{one}-dimensional SPT with respect to the residual symmetry group $H$, corresponding to a choice of an element of $\mathcal{H}^2(H, \mathrm{U}(1))$. In summary, the possible line-like excitations carrying flux $x \in G$ are specified by the following data:
\begin{framed}
\textbf{Data for loop-like excitations carrying flux $x$.}
\begin{itemize}
\item A subgroup $H \leq G$ such that $(\widetilde{\omega}_x)_H$ is a trivial 3-cocycle of $H$.
\item An element $\gamma \in \mathcal{H}^2(H, \mathrm{U}(1))$.
\end{itemize}
\end{framed}
The second piece of data should more precisely be an element of a torsor over $\mathcal{H}^2(H, \mathrm{U}(1))$. This is because the identification of the (1+1)-D SPT carried on the boundary depends on a choice of disentangler to separate the boundary from the bulk, and different disentanglers could differ by a pumping of an (1+1)-D SPT onto the boundary, so it isn't possible to identify when the boundary carries ``trivial'' SPT.
In fact, the only physical manifestation of this feature of loop excitations is the presence on a loop of pointlike defects
carrying projective representations of $H$; they occur at the boundary between different gauged
(1+1)-D SPTs that occur on different arc-like sections of a loop excitation.

Let us remark that $H$, though defined in terms of the SPT boundary, has a number of physical manifestations for the original loop excitation in the (3+1)-D topological phase. Firstly, it determines the group $A$ of possible Cheshire charges on the loop (the ``condensation group''). Recall that, in general, a charge in a gauge theory with Abelian gauge group is labelled by a one-dimensional representation $\chi \in G^{*}$. By considerations similar to those in the previous section, one finds that in general the topological degeneracy corresponds to the Cheshire charge ranging over $A = \{ \chi \in G^{*} : \chi(h) = 1 \quad \forall h \in H\}$. Thus, the topological degeneracy is $|A| = |G|/|H|$, where $|\cdot|$ denotes the size of a group. This is seen to be in agreement with the number of symmetry-breaking ground states in the SPT boundary. (However, it corresponds to a different choice of basis.) From the SPT boundary picture one also sees that that $A$ also describes the charges which are allowed to locally annihilate on the loop, justifying the name ``condensation group''.

Another interpretation of $H$ is as the set of fluxes that are allowed to pass through the given loop excitation. Indeed, applying the general constraint that is boxed in Sec.~\ref{sec:cheshire-charge-general} shows that two loop excitations  $(x,H,\gamma)$ and $(x',H',\gamma')$ are obstructed from passing through each other except when $x \in H'$ and $x' \in H$.

\subsection{Example: $G = Z_2 \times Z_2$}
\tikzexternaldisable

As an example, let us consider the case of $G = \mathbb{Z}_2 \times \mathbb{Z}_2$. This is the simplest gauge group $G$ for which a non-trivial 4-cocycle exists; one of them is
\begin{equation}
\label{wZ}
\omega( g_1, g_2, g_3, g_4 ) = (-1)^{g_1^{(0)} g_2^{(1)} g_3^{(1)} g_4^{(1)} + g_1^{(1)} g_2^{(0)} g_3^{(0)} g_4^{(0)}}
\end{equation}
where we write the elements of $G = \mathbb{Z}_2 \times \mathbb{Z}_2$ as $g = (g^{(0)}, g^{(1)})$ with $g^{(0)}, g^{(1)} \in \{0, 1\}$. We show in Table \ref{Z2Z2_table} the dimensionally reduced 3-cocycles $\widetilde{\omega}_x$ and the allowed unbroken subgroups $H \leq G$.
\begin{table*}
\begin{tabular}{|c|c|c|c|c|c|c|c|c}
\hline
\multirow{2}{*}{$x$} &
\multirow{2}{*}{$\widetilde{\omega}_x(g_1, g_2, g_3)$}
                       & \multicolumn{5}{|c|}{Allowed unbroken subgroups $H$} \\
    \cline{3-7}
  &  & 1 & $\langle (1,0) \rangle$ & $\langle (0,1) \rangle$ & $\langle (1,1) \rangle$ & $\mathbb{Z}_2 \times \mathbb{Z}_2$ \\
\hline
(0,0) & 1 & \cmark & \cmark & \cmark & \cmark & \cmark \\
(1,0) & $(-1)^{g_1^{(1)} ( g_2^{(1)} g_3^{(1)} + g_2^{(0)} g_3^{(0)})}$  & \cmark & \cmark & \xmark & \cmark & \xmark \\
(0,1) & $(-1)^{g_1^{(0)}( g_2^{(1)} g_3^{(1)} + g_2^{(0)} g_3^{(0)} )}$  & \cmark & \xmark & \cmark & \cmark & \xmark \\
(1,1) & $(-1)^{(g_1^{(0)} + g_1^{(1)})( g_2^{(1)} g_3^{(1)} + g_2^{(0)} g_3^{(0)} )}$ & \cmark & \xmark & \xmark & \cmark & \xmark \\
\hline
\end{tabular}
\caption{\label{Z2Z2_table}The dimensionally reduced 3-cocycles and allowed residual (un-Higgsed) gauge groups for loop excitations in the (3+1)-D Dijkgraaf-Witten theory with gauge group $G = \mathbb{Z}_2 \times \mathbb{Z}_2$ and 4-cocycle \eqnref{wZ}. Here $\langle g \rangle$ denotes the subgroup generated by the group element $g$.}
\end{table*}

\section{Constructing the loop excitations explicitly in a (3+1)-D lattice model}
\label{sec:lattice-model}

Although the dimensional reduction arguments of the previous two sections are sufficient to establish the existence of the Cheshire charge phenomemon and the classification of the loop excitations, it is instructive to construct the loop excitations explicitly in a fully (3+1)-D system. We do this though a slight extension of the work of Ref.~\onlinecite{Lin15}, in which an exactly solvable model of a (3+1)-D twisted $\mathbb{Z}_2 \times \mathbb{Z}_2$ gauge theory was constructed. We believe that the topological order of this model is precisely the Dijkgraaf-Witten theory with 4-cocycle \eqnref{wZ}.
Other lattice models realizing such phases were constructed in Refs.\onlinecite{Walker12,Wan15,Williamson16,Wang16}; we believe that an enumeration
of the loop excitations of these models will give similar results, but we do not undertake
this here.

We will briefly sketch the construction of Ref.~\onlinecite{Lin15}, referring the reader to that paper for further details. The model has a spin-1/2 particle (``red spins'') on each face of some three-dimensional lattice, and another (``blue'') on each face of the dual lattice. We imagine that the state of each red spin tells us about the presence or absence of a ``red membrane'' on the corresponding face, and similarly for the blue spins. Thus, a basis for the Hilbert space is labelled by configurations of red and blue membranes living on the lattice and its dual respectively. Recall that we expect flux lines in a $\mathbb{Z}_2 \times \mathbb{Z}_2$ gauge theory to be labelled by $\mathbb{Z}_2 \times \mathbb{Z}_2$, which has two generators (hence, two ``elementary'' flux types). In the membrane representation, these correspond to boundaries of red and blue membranes, respectively.

The ground state wavefunction for the twisted gauge theory is
\begin{equation}
\ket{\Psi_1} = \sum_{X_r,X_b} (-1)^{N(X_r, X_b)} \ket{X_r,X_b}. \label{twisted_wf}
\end{equation}
where the sum is over all configurations of \emph{closed} red and blue membranes, and $N= (X_r, X_b)$ is the number of intersection loops between red and blue membranes. One can contrast this with the wavefunction
\begin{equation}
\ket{\Psi_0} = \sum_{X_r,X_b} \ket{X_r,X_b}, \label{untwisted_wf}
\end{equation}
(where the sum is again over closed membrane configurations), which represents the ground state of an \emph{untwisted} $\mathbb{Z}_2 \times \mathbb{Z}_2$ gauge theory.

In order to construct loop-like excitations for the wavefunctions \eqnref{twisted_wf} and \eqnref{untwisted_wf}, the authors of Ref.~\onlinecite{Lin15} first go through the intermediate step of constructing ``membrane operators'' ${M_b}(S)$, ${M_r}(S)$ which act on closed surfaces $S$ in three-dimensional space and leave the ground state invariant. If one then restricts such an operator in an appropriate way to a surface with boundary and applies it the ground state, one creates some excitation which necessarily differs from the ground state only on the boundary. Let us focus, for the
sake of concreteness, on the membrane operators that create blue flux excitations upon restriction to a surface with boundary. These take the form
\begin{equation}
{M_b}(S) = \sum_{X_r, X_b} {f_b}({X_r}, {X_b}, S) \ket{X_r, X_b + S} \bra{X_r, X_b},
\end{equation}
where we take the surface $S$ to be made out of blue faces, and
where $X_b + S$ denotes the configuration obtained from $X_b$ by flipping the spin at each blue face in $S$. The coefficient ${f_b}$ depends only on the intersections of blue and red membranes with $S$ and is determined by a set of implicit rules motivated by the requirement that ${M_b}(S)$ leave the ground state invariant.

These rules can be represented graphically. Intersections between $S$ and the red membranes in $X_r$
are depicted by red lines. As the membrane $S$ is defined on the same faces as the blue membranes, there can be regions where $S$ is coincident with a blue membrane in $X_b$. The \emph{boundary} of such a region is depicted by a blue line, which is either dotted or solid according to whether the blue membrane in
$X_b$ is incident on $S$ either from above or below, which is defined with respect to a choice of
normal vector to $S$ (see Ref.~\onlinecite{Lin15} for a more detailed explanation).
The rules can be written as:
\begin{subequations}
\label{eqn:local}
\begin{align}
\label{eqn:red-rule1}
{f_b}\left(
\begin{tikzpicture}[baseline=-0.1 cm] \draw[red,thick] (0,0) circle (10 pt); \end{tikzpicture}
\right) &= \sigma {f_b}(\cdot) \, ,
\\
\label{eqn:red-rule2}
{f_b}\left(
\begin{tikzpicture}[baseline=0.25 cm] \draw[red,thick] (0,0) ..controls (0.35,0.23) and (0.35,0.47) .. (0,0.7) ;
\draw[red,thick] (0.75,0) ..controls (0.35,0.23) and (0.35,0.47) .. (0.75,0.7);
 \end{tikzpicture}
\right) &= \sigma
{f_b}\left(\begin{tikzpicture}[baseline=0.25 cm] \draw[red,thick] (0,0) ..controls (0.25,0.35) and (0.5,0.35) .. (0.75,0);
\draw[red,thick] (0,0.7) ..controls (0.25,0.35) and (0.5,0.35) .. (0.75,0.7);
 \end{tikzpicture}\right) \, ,
\\
\label{eqn:blue-rule1}
{f_b}\left(
\begin{tikzpicture}[baseline=-0.1 cm] \draw[blue,thick] (0,0) circle (10 pt); \end{tikzpicture}
\right) &= f(\cdot) \, ,
\\
\label{eqn:blue-rule2}
{f_b}\left(
\begin{tikzpicture}[baseline=0.25 cm] \draw[blue,thick] (0,0) ..controls (0.35,0.23) and (0.35,0.47) .. (0,0.7) ;
\draw[blue,thick] (0.75,0) ..controls (0.35,0.23) and (0.35,0.47) .. (0.75,0.7);
 \end{tikzpicture}
\right) &=
{f_b}\left(\begin{tikzpicture}[baseline=0.25 cm] \draw[blue,thick] (0,0) ..controls (0.25,0.35) and (0.5,0.35) .. (0.75,0);
\draw[blue,thick] (0,0.7) ..controls (0.25,0.35) and (0.5,0.35) .. (0.75,0.7);
 \end{tikzpicture}\right) \, ,
\\
\label{eqn:blue-rule3}
{f_b}\left(
\begin{tikzpicture}[baseline=-0.1 cm] \draw[blue,thick,dotted] (0,0) circle (10 pt); \end{tikzpicture}
\right) &= f(\cdot) \, ,
\\
\label{eqn:blue-rule4}
{f_b}\left(
\begin{tikzpicture}[baseline=0.25 cm] \draw[blue,thick,dotted] (0,0) ..controls (0.35,0.23) and (0.35,0.47) .. (0,0.7) ;
\draw[blue,thick,dotted] (0.75,0) ..controls (0.35,0.23) and (0.35,0.47) .. (0.75,0.7);
 \end{tikzpicture}
\right) &=
{f_b}\left(\begin{tikzpicture}[baseline=0.25 cm] \draw[blue,thick,dotted] (0,0) ..controls (0.25,0.35) and (0.5,0.35) .. (0.75,0);
\draw[blue,thick,dotted] (0,0.7) ..controls (0.25,0.35) and (0.5,0.35) .. (0.75,0.7);
 \end{tikzpicture}\right) \, ,
\\
\label{eqn:blue-rule5}
{f_b}\left(
\begin{tikzpicture}[baseline=0.25 cm] \draw[blue,thick] (0,0) ..controls (0.8,0.23) and (0.8,0.47) .. (0,0.7) ;
\draw[blue,thick,dotted] (0.75,0) ..controls (-0.1,0.23) and (-0.1,0.47) .. (0.75,0.7);
 \end{tikzpicture}
\right) &=
{f_b}\left(
\begin{tikzpicture}[baseline=0.25 cm] \draw[blue,thick] (0,0) ..controls (0.35,0.23) and (0.35,0.47) .. (0,0.7) ;
\draw[blue,thick,dotted] (0.75,0) ..controls (0.35,0.23) and (0.35,0.47) .. (0.75,0.7);
 \end{tikzpicture}
\right) \, ,
\\
\label{eqn:blue-red-rule1}
{f_b}\left(
\begin{tikzpicture}[baseline=0.25 cm] \draw[red,thick] (0,0) ..controls (0.8,0.23) and (0.8,0.47) .. (0,0.7) ;
\draw[blue,thick] (0.75,0) ..controls (-0.1,0.23) and (-0.1,0.47) .. (0.75,0.7);
 \end{tikzpicture}
\right) &=
{f_b}\left(
\begin{tikzpicture}[baseline=0.25 cm] \draw[red,thick] (0,0) ..controls (0.35,0.23) and (0.35,0.47) .. (0,0.7) ;
\draw[blue,thick] (0.75,0) ..controls (0.35,0.23) and (0.35,0.47) .. (0.75,0.7);
 \end{tikzpicture}
\right) \, ,
\\
\label{eqn:blue-red-rule2}
{f_b}\left(
\begin{tikzpicture}[baseline=0.25 cm] \draw[red,thick] (0,0) ..controls (0.8,0.23) and (0.8,0.47) .. (0,0.7) ;
\draw[blue,thick,dotted] (0.75,0) ..controls (-0.1,0.23) and (-0.1,0.47) .. (0.75,0.7);
 \end{tikzpicture}
\right) &=
{f_b}\left(
\begin{tikzpicture}[baseline=0.25 cm] \draw[red,thick] (0,0) ..controls (0.35,0.23) and (0.35,0.47) .. (0,0.7) ;
\draw[blue,thick,dotted] (0.75,0) ..controls (0.35,0.23) and (0.35,0.47) .. (0.75,0.7);
 \end{tikzpicture}
\right) \, ,
\\
\label{eqn:blue-red-rule3}
{f_b}\left(
\begin{tikzpicture}[baseline=0.25 cm] \draw[red,thick] (0.17,0.75) -- (0.3,0) ;
\draw[blue,thick] (0,0.7) -- (0.75,0);
\draw[blue,thick,dotted] (0,0) -- (0.75,0.7);
 \end{tikzpicture}
\right) &= \sigma
{f_b}\left(
\begin{tikzpicture}[baseline=0.25 cm] \draw[red,thick] (0.15,0.75) ..controls (0.75,0.45) and (0.75,0.25) .. (0.3,0) ;
\draw[blue,thick] (0,0.7) -- (0.75,0);
\draw[blue,thick,dotted] (0,0) -- (0.75,0.7);
 \end{tikzpicture}
\right) \, ,
\end{align}
\end{subequations}
where $\sigma$ is $+1$ or $-1$ in the untwisted and twisted cases, respectively.
We have not explicitly depicted the blue shading in these rules. It can be added in any way that is consistent
on both sides of the equation (so that each rule involving blue lines will be, strictly speaking, multiple rules,
one for each possible consistent shading).

Now, to actually construct the loop excitations, we need to define the operators ${M_b}(S)$
on surfaces $S$ with boundary. In that case we need to supplement the rules in \eqnref{eqn:local} by additional conditions at the boundary. The simplest possible choice, made by Ref.~\onlinecite{Lin15} in the
untwisted case, is simply to allow allow both red and blue membranes to intersect the boundary of $S$
or, in other words, to simply set ${M_b}(S)=1$.
This choice gives an operator that creates a particular type of loop excitations which, as we will see,
do not carry Cheshire charge. Charge can be added to such a loop with a string operator \cite{Lin15}
that creates charged excitations at its endpoints: if one of the endpoints lies on a flux loop, then the loop carries charge. However, there is an additional energy cost associated with this charge, namely the energy cost of a string endpoint,
so there is no topological degeneracy associated with the presence or absence of such charge.
However, there are other possible choices of ${M_b}(S)$, and we will find that making different choices of boundary conditions allows us to reproduce the various kinds of excitations that exist in a given flux sector.
For instance, in the twisted case the operator ${M_b}(S)$ was chosen in Ref.~\onlinecite{Lin15} to
project onto the states in which no red or blue membrane intersects the boundary of $S$.
In this case, a blue loop can carry either blue or red Cheshire charge, and the loop has fourfold
degeneracy. The existence of four different blue loop excitations was noted in Table I of
Ref.~\onlinecite{Lin15}; the new point that we are emphasizing here is that these four loops excitations
\emph{cannot be locally distinguished} and, therefore, must be degenerate in energy.

We now construct the different types of loop excitations more systematically.
We first note that we have four choices: we can choose to project onto the states in which no membranes intersect the boundary; in which only \emph{red} membranes intersect the boundary; in which only \emph{blue} membranes intersect the boundary; or perform no projection at all.
We then require $f$ to satisfy the following rules at the boundary, which is represented by a thick gray line:
\begin{subequations}
\label{eqn:boundary-rules}
\begin{align}
\label{eqn:red-boundary-rule1}
{f_b}\!\left(
\begin{tikzpicture}[baseline=0.2 cm] \draw[gray,very thick] (0,0) -- (1,0);
\draw[red,thick] (0.2,0) ..controls (0.2,0.7) and (0.8,0.7) .. (0.8,0);
\end{tikzpicture}\right)
&= {\alpha_r} {f_b}\!\left(
\begin{tikzpicture}[baseline=0.2 cm] \draw[gray,very thick] (0,0) -- (0.9,0);
\end{tikzpicture}\right) \, ,
\\
\label{eqn:blue-boundary-rule1}
{f_b}\!\left(
\begin{tikzpicture}[baseline=0.2 cm] \draw[gray,very thick] (0,0) -- (1,0);
\draw[blue,thick] (0.2,0) ..controls (0.2,0.7) and (0.8,0.7) .. (0.8,0);
\end{tikzpicture}\right)
&= {\alpha_b} {f_b}\!\left(
\begin{tikzpicture}[baseline=0.2 cm] \draw[gray,very thick] (0,0) -- (0.9,0);
\end{tikzpicture}\right) \, ,
\\
\label{eqn:blue-boundary-rule2}
{f_b}\!\left(
\begin{tikzpicture}[baseline=0.2 cm] \draw[gray,very thick] (0,0) -- (1,0);
\draw[blue,thick,dotted] (0.2,0) ..controls (0.2,0.7) and (0.8,0.7) .. (0.8,0);
\end{tikzpicture}\right)
&= {{\tilde \alpha}_b} {f_b}\!\left(
\begin{tikzpicture}[baseline=0.2 cm] \draw[gray,very thick] (0,0) -- (0.9,0);
\end{tikzpicture}\right) \, ,
\\
\label{eqn:blue-red-boundary-rule1}
{f_b}\!\left(
\begin{tikzpicture}[baseline=0.2 cm] \draw[gray,very thick] (0,0) -- (1,0);
\draw[blue,thick] (0.2,0) ..controls (0.2,0.7) and (0.8,0.7) .. (0.8,0);
\draw[red,thick] (0.5,0) -- (0.5,0.7);
\end{tikzpicture}\right)
&= {\beta} {f_b}\!\left(
\begin{tikzpicture}[baseline=0.2 cm] \draw[gray,very thick] (0,0) -- (0.9,0);
\draw[red,thick] (0.45,0) -- (0.45,0.7);
\end{tikzpicture}\right) \, ,
\\
\label{eqn:blue-red-boundary-rule2}
{f_b}\!\left(
\begin{tikzpicture}[baseline=0.2 cm] \draw[gray,very thick] (0,0) -- (1,0);
\draw[blue,thick,dotted] (0.2,0) ..controls (0.2,0.7) and (0.8,0.7) .. (0.8,0);
\draw[red,thick] (0.5,0) -- (0.5,0.7);
\end{tikzpicture}\right)
&= {\tilde \beta} {f_b}\!\left(
\begin{tikzpicture}[baseline=0.2 cm] \draw[gray,very thick] (0,0) -- (0.9,0);
\draw[red,thick] (0.45,0) -- (0.45,0.7);
\end{tikzpicture}\right) \, ,
\end{align}
\end{subequations}
for some phase factors $\alpha_r, \alpha_b, \widetilde{\alpha}_b, {\beta}, \widetilde{\beta}$.
The phase factors $\alpha_r, \beta, \widetilde{\beta}$ are only required for those membrane operators
that do not annihilate states in which a red membrane intersects the boundary. (If such states \emph{are} annihilated, then they are effectively set to zero.) Similarly, $\alpha_b, \widetilde{\alpha_b}, \beta, \widetilde{\beta}$ are only
required for those membrane operators that do not annihilate states in which
a blue membrane overlaps the boundary.
These coefficients must satisfy consistency conditions in order to be compatible with the rules \eqnref{eqn:local} away from the boundary. Indeed, we have 
\begin{align}
\label{eqn:boundary-consistency}
{f_b}\!\left(
\begin{tikzpicture}[baseline=0.2 cm] \draw[gray,very thick] (0,0) -- (1,0);
\draw[blue,thick] (0.2,0) ..controls (0.2,0.7) and (0.8,0.7) .. (0.8,0);
\draw[blue,thick] (0.35,0) ..controls (0.35,0.55) and (0.65,0.55) .. (0.65,0);
\draw[red,thick] (0.5,0) -- (0.5,0.7);
\end{tikzpicture}\right)
&= {f_b}\!\left(
\begin{tikzpicture}[baseline=0.2 cm] \draw[gray,very thick] (0,0) -- (1,0);
\draw[blue,thick] (0.2,0) ..controls (0.2,0.4) and (0.4,0.4) .. (0.35,0);
\draw[blue,thick] (0.65,0) to [out=90, in=350] (0.47,0.55);
\draw[blue,thick] (0.47,0.55) to [out=170, in=0] (0.3,0.4);
\draw[blue,thick] (0.3,0.4) to [out=180, in=240] (0.3,0.6);
\draw[blue,thick] (0.3,0.6) to [out=60, in=180] (0.5,0.7);
\draw[blue,thick] (0.5,0.7) to [out=0, in=90] (0.8,0);
\draw[red,thick] (0.5,0) -- (0.5,0.7);
\end{tikzpicture}\right) \nonumber\\
& = \sigma \, {f_b}\!\left(
\begin{tikzpicture}[baseline=0.2 cm] \draw[gray,very thick] (0,0) -- (1,0);
\draw[blue,thick] (0.2,0) ..controls (0.2,0.5) and (0.4,0.5) .. (0.35,0);
\draw[blue,thick] (0.65,0) ..controls (0.6,0.5) and (0.80,0.5) .. (0.8,0);
\draw[red,thick] (0.5,0) -- (0.5,0.7);
\end{tikzpicture}\right) \nonumber\\
& = \sigma {\alpha_b^2}\, {f_b}\!\left(
\begin{tikzpicture}[baseline=0.2 cm] \draw[gray,very thick] (0,0) -- (1,0);
\draw[red,thick] (0.5,0) -- (0.5,0.7);
\end{tikzpicture}\right)\nonumber\\
& = \sigma \beta_b^{-1} {\alpha_b^2}\, {f_b}\!\left(
\begin{tikzpicture}[baseline=0.2 cm] \draw[gray,very thick] (0,0) -- (1,0);
\draw[blue,thick] (0.2,0) ..controls (0.2,0.7) and (0.8,0.7) .. (0.8,0);
\draw[red,thick] (0.5,0) -- (0.5,0.7);
\end{tikzpicture}\right) \nonumber\\
& = \sigma \beta_b^{-2} {\alpha_b^2}\, {f_b}\!\left(
\begin{tikzpicture}[baseline=0.2 cm] \draw[gray,very thick] (0,0) -- (1,0);
\draw[blue,thick] (0.2,0) ..controls (0.2,0.7) and (0.8,0.7) .. (0.8,0);
\draw[blue,thick] (0.35,0) ..controls (0.35,0.55) and (0.65,0.55) .. (0.65,0);
\draw[red,thick] (0.5,0) -- (0.5,0.7);
\end{tikzpicture}\right)
\end{align}
from which we conclude that $\sigma {\alpha_b^2} = \beta^{2}$.
By similar logic, we conclude that $\sigma \widetilde{\alpha}_b^2 = \widetilde{\beta}^2$
and we also obtain the condition:
\begin{align}
{\alpha^2_r}\,\, {f_b}\!\left(
\begin{tikzpicture}[baseline=0.2 cm] \draw[gray,very thick] (0,0) -- (0.9,0);
\end{tikzpicture}\right) &=
{f_b}\!\left(
\begin{tikzpicture}[baseline=0.2 cm] \draw[gray,very thick] (0,0) -- (1,0);
\draw[red,thick] (0.2,0) ..controls (0.2,0.7) and (0.8,0.7) .. (0.8,0);
\draw[red,thick] (0.35,0) ..controls (0.35,0.55) and (0.65,0.55) .. (0.65,0);
\end{tikzpicture}\right) \nonumber\\
& = \sigma \, {f_b}\!\left(
\begin{tikzpicture}[baseline=0.2 cm] \draw[gray,very thick] (0,0) -- (1,0);
\draw[red,thick] (0.2,0) ..controls (0.2,0.5) and (0.4,0.5) .. (0.35,0);
\draw[red,thick] (0.65,0) ..controls (0.6,0.5) and (0.80,0.5) .. (0.8,0);
\end{tikzpicture}\right) \nonumber\\
& = \sigma {\alpha^2_r}\, {f_b}\!\left(
\begin{tikzpicture}[baseline=0.2 cm] \draw[gray,very thick] (0,0) -- (0.9,0);
\end{tikzpicture}\right)
\end{align}
The resulting constraint ${\alpha^2_r} = \sigma {\alpha^2_r}$
implies a contradiction in the twisted case ($\sigma = -1$), unless $\alpha_r = 0$. Therefore, in the twisted case,
the membrane operator \emph{must} annihilate the states in which a red membrane intersects the boundary.

Two different choices of the coefficients in \eqnref{eqn:boundary-rules}
do not necessarily produce different excitations. Indeed, suppose we act on the boundary with a unitary 
\begin{multline}
U = 
\sum_{N_r, N_b} \gamma_r^{K_r(N_r)} \widetilde{\gamma_r}^{\widetilde{K_r}(N_r)} \gamma_b^{K_b(N_b)}\,\times \\ \widetilde{\gamma_b} ^{\widetilde{K_b}(N_b)} \ket{N_r, N_b} \bra{N_r, N_b},
\end{multline}
where $K_r$ and $\widetilde{K}_r$ are the number of red lines intersecting the boundary in an unshaded and shaded region, respectively (recall that blue lines, whether solid or dotted, enclose shaded regions); $K_b$ and $\widetilde{K}_b$ are the number of solid and dashed blue lines intersecting the boundary respectively; and $\gamma_r$, $\widetilde{\gamma_r}$, $\gamma_b$, $\widetilde{\gamma_b}$ are phase factors. We take $\gamma_r = \lambda \widetilde{\gamma_r}$ for some $\lambda = \pm 1$ in order to ensure consistency with the rules
(\ref{eqn:local}) and (\ref{eqn:boundary-rules}), which hold for all possible shadings of the pictures.
This corresponds to changing
\begin{subequations}
\label{boundary_gauge}
\begin{align}
\alpha_r &\to \alpha_r \gamma_r^2 \\
\alpha_b &\to \alpha_b \gamma_b^2 \\
\widetilde{\alpha_b} &\to \widetilde{\alpha_b} \widetilde{\gamma_b}^2\\
\beta &\to \lambda \beta \\ 
\widetilde{\beta} &\to \lambda \widetilde{\beta}
\end{align}
\end{subequations}
On the other hand, since $U$ is a local unitary it is clear that we should identify the corresponding loop excitations as being essentially the same. Thus, the different loop excitations are characterized by the choice of projection and an \emph{equivalence class} of the boundary data under the transformation \eqnref{boundary_gauge}. In fact, if we require Eq.~(\ref{eqn:boundary-consistency}) and the analogous equation for $\widetilde{\beta}$ to be satisfied, there are at most two distinct equivalence classes under this transformation. In the case where the membrane operator does not annihilate any states, there are two equivalence classes correpsonding to the relative sign of $\beta$ and $\widetilde{\beta}$; in all other cases, $\beta$ and $\widetilde{\beta}$ are effectively set to zero and there is only one equivalence class.

We now observe that, depending on the choice of projection procedure, the aforementioned rules might still not determine the membrane operator uniquely in the case where the surface $S$ is a half-infinite cylinder, as was observed in Ref.~\onlinecite{Lin15}.
Suppose that the membrane operator annihilates the states in which a red membrane intersects the boundary (this must be the case in the twisted theory; it may or may not be the case in the untwisted
theory since both types of membrane operators are allowed).
Then there is a well-defined $\mathbb{Z}_2$ \emph{winding number} for red loops around the cylinder, which is unaffected by any of the local moves \eqnref{eqn:local}. Thus, the amplitude for the two winding number sectors must be specified separately. There is thus a \emph{two-dimensional} space of states that can be created, depending on this choice of amplitudes. Furthermore, since these states differ only by a winding number, which is not a locally measurable quantity, these states are locally indistinguishable. It can be verified that a basis for these states can be found which differs in the overall charge on the boundary of the cylinder. Thus, we interpret this degeneracy as corresponding to Cheshire charge. (On the other hand, if we do not project out the states in which a red membrane intersects the boundary, and instead impose
rules (\ref{eqn:red-boundary-rule1}), (\ref{eqn:blue-red-boundary-rule1}), and (\ref{eqn:blue-red-boundary-rule2}), then the winding number is not well-defined and there is no associated degeneracy.) Similarly, if we have projected out all states in which a \emph{blue} membrane intersects the boundary, there is also another $\mathbb{Z}_2$ winding number for blue loops, with associated degeneracy.

\begin{table}
\begin{tabular}{|c|c|c|c|c|}
\hline
\multicolumn{2}{|c|}{Project out intersections} & \multirow{2}{*}{Degeneracy} & \multirow{2}{*}{Residual group}  \\
\cline{1-2}
Red & Blue &  & \\
\hline
Yes & Yes & 4 & 1 \\
Yes & No & 2 & $\langle (1,0) \rangle$ \\
\rowcolor{Gray}[\tabcolsep][\tabcolsep]
No & Yes & 2 & $\langle (0,1) \rangle$ \\
\rowcolor{Gray}[\tabcolsep][\tabcolsep]
No & No & 1 & $\mathbb{Z}_2 \times \mathbb{Z}_2$ \\
\hline
\end{tabular}
\caption{\label{excitation_types}The excitation types for which we have constructed membrane operators in the exactly-solvable $\mathbb{Z}_2 \times \mathbb{Z}_2$ gauge theory. In the untwisted case ($\sigma=1$) all four rows are allowed, whereas in the twisted case ($\sigma=-1$) only the first two rows are allowed.}
\end{table}
In summary, therefore, we have found the excitation types listed in Table \ref{excitation_types},
where in the last column we have identified the residual gauge group on the excitation in accordance with the dimensional reduction treatment of Sec. \ref{dimreduction_enumerate}. As we have seen, the last row actually includes two different excitation types (determined by the relative sign of $\beta$ and $\widetilde{\beta}$); this corresponds to the second piece of data discussed in Section \ref{dimreduction_enumerate}, because $\mathcal{H}^2(\mathbb{Z}_2 \times \mathbb{Z}_2, \mathrm{U}(1)) = \mathbb{Z}_2$. Comparing with Table \ref{Z2Z2_table}, we see that we have achived perfect agreement with the classication of Section \ref{dimreduction_enumerate} for the case of a $(1,0)$ flux, except that we have not reproduced the excitations with residual group $\langle (1,1) \rangle$; we expect that they could also be constructed if we allowed for a more complicated choice of boundary condition.
\tikzexternalenable

\section{More general constructions: membrane-net models}
The exactly solvable model considered above is rather peculiar to the case of a $\mathbb{Z}_2 \times \mathbb{Z}_2$ gauge group. A more general approach is through the theory of ``membrane-net condensation''. These are the natural extensions of the ``string-net condensation'' of Levin and Wen \onlinecite{Levin05a} to 3D, and likewise give rise to exactly solvable models of topological phases. In Appendix \ref{membranenet}, we outline how the ``membrane operators'' creating loop excitations in the membrane-net models that correspond to Dijkgraaf-Witten theory indeed obey the classification of Section \ref{dimreduction_enumerate}.

\section{Category-theoretical viewpoint}
\label{sec:2-categories}

In Refs.~\onlinecite{Kong2014,Kong2015}, it has been conjectured that a branch of mathematics known as higher category theory provides the natural language to describe excitations in $n$-dimensional topological phases (see also Ref.~\onlinecite{Baez1995} for an introduction to higher category theory). In paticular, the excitations in a topological phase in $n$ spatial dimensions ought to be describable by a mathematical object known as a unitary braided fusion $(n-1)$-category. Such a statement necessarily must remain as a conjecture, since even the correct definition of an $n$-category is still a subject of current research in the mathematics community.

Nevertheless, the results of this paper provide some evidence in favor of this conjecture. Specifically, we want to show that the loop-like excitations classified in Section \ref{dimreduction_enumerate} for a (3+1)-D Dijkgraaf Witten theory with Abelian  gauge group $G$ and 4-cocycle $\omega$ can be realized as objects in a particular braided fusion $2$-category. (The objects of a braided fusion 2-category are only a small part of the whole mathematical structure; we leave for future work the question of how to relate all the remaining structure to the physical properties of the loop excitations.)  It remains to identify the appropriate braided fusion 2-category.

To this end, we note that the original string-net construction of Levin-Wen \cite{Levin05a}, and its membrane-net version in (3+1)-D discussed in Appendix \ref{membranenet}, has a strong categorical flavor. The input to a Levin-Wen model is essentially a fusion category $\mathcal{C}$, and the string operators used by Levin and Wen to construct excitations correspond to objects of the \emph{monoidal center} $Z(\mathcal{C})$. Similarly, the input to a membrane-net construction along the lines of Appendix \ref{membranenet} is essentially a fusion 2-category $\mathcal{C}$, and the process of constructing the membrane operators that create excitations, as described in Appendix \ref{membranenet}, \emph{exactly} corresponds to finding the objects in the monoidal center $Z(\mathcal{C})$.
The particular fusion 2-category corresponding to a Dijkgraaf-Witten model with finite Abelian  gauge group $G$ and 4-cocycle $\omega$ turns out to be $\mathcal{C} = \cat{2Vect}_G^{\omega}$, the category of $G$-graded 2-vector spaces, twisted by $\omega$ (see Appendix \ref{sec:def-2Vect} for a definition.) Therefore, the excitations in such a theory correspond to objects in $Z(\cat{2Vect}_G^{\omega})$.

%

\begin{acknowledgments}
We thank C.-H. Lin, M. Cheng, J. Rickard, J. Wang and M. Barkeshli for helpful discussions. DVE is supported by the Microsoft Corporation.
\end{acknowledgments}

\appendix

\section{The flux-(SPT boundary) correspondence in a (3+1)-D Dijkgraaf-Witten theory}
\label{appendix_spt}
In this section, we specialize to Dijkgraaf-Witten topological phases in (3+1)-D, characterized by a 4-cocycle $\omega(g_1, g_2, g_3, g_4)$. We can understand the structure of loop excitations in this theory by exploiting the connection between Dijkgraaf-Witten topologicl phases and symmetry-protected topological (SPT) phases. The former are obtained by gauging the latter. The resulting Dijkgraaf-Witten theory is described by the same 4-cocycle $\omega(g_1, g_2, g_3, g_4)$ as the original SPT phase.
The gauging process can be thought of as increasing the degrees of freedom of the system by coupling to a new field -- the gauge field. Specifically, let $\mathcal{H}$ be the Hilbert space of the bosonic SPT system. We introduce an enlarged Hilbert space $\mathcal{H}'$ to describe configurations of the gauge field as well as the original degrees of freedom. Specifically, we have $\mathcal{H}' = \mathcal{H} \otimes \mathcal{H}_{\mathrm{gauge}}$, where $\mathcal{H}_{\mathrm{gauge}}$ is spanned by a basis $| \{ \mu_{i,j} \} \rangle$, where $\mu_{i,j}$ assigns a group element $g \in G$ for every link $\langle i, j\rangle$ in the lattice. (We are also supposed to identify states in $\mathcal{H'}$ related by gauge transformations, or equivalently, project onto the gauge-invariant subspace of $\mathcal{H}'$.) Let $|0\rangle_{\mathrm{gauge}}$ be the state corresponding to all the $\mu_{i,j} = 1$. The gauged Hamiltonian $H'$ is chosen such that it reduces to $H$ on the subspace $\mathcal{H} \otimes |0\rangle_{\mathrm{gauge}}$.

We will henceforth consider a Hamiltonian for the Dijkgraaf-Witten theory such that the gauge curvature is not allowed to fluctuate; that is, the ground state is constrained to be a superposition only over flat gauge configurations. This is true, for example, for the model of Ref.~\onlinecite{Lin15}),
and in general can be done for any Dijkgraaf-Witten theory. (Moreover, the ground state of a Dijkgraaf-Witten theory that does not satisfy  such a constraint can be related to one which does by a local unitary, so the below arguments still carry over.) It follows that, on a simply connected space, the ground state can be written
can be written as the gauge-invariant projection of $\ket{\Psi}_{\mathrm{SPT}} \otimes \ket{0}_{\mathrm{gauge}}$, where $\ket{\Psi}_{\mathrm{SPT}}$ is the SPT ground state.

Now consider a loop-like excitation in the Dijkgraaf-Witten phase. In general, the excitation can carry gauge flux, in which case it does not correspond to a loop-like excitation of a bosonic SPT. However, since the gauge field is still required to be flat away from the loop, it follows that we can make a gauge transformation to trivialize the gauge field on a simply connected region $R$ enclosing the loop. (Specifically, we will take $R$ to be the complement of a disc-like ``pancake'' enclosing the loop) In other words, the loop excitation state can be expressed as the gauge-invariant projection of a state $\ket{\Psi} _{\mathrm{loop}} = \ket{\widetilde{\Psi}}_{\mathrm{loop}} \otimes \ket{0}_{\mathrm{gauge,R}}$. Here we have employed a different tensor product decomposition from before: $\mathcal{H}' = \widetilde{\mathcal{H}} \otimes \mathcal{H}_{R,\mathrm{gauge}}$ where $\mathcal{H}_{\mathrm{R,gauge}}$ corresponds to the configurations of the gauge field within $R$ only. Without loss of generality, we can take $\ket{\widetilde{\Psi}}$ to be invariant under the global part of the gauge symmetry, whose action we write as $U(g)$. Therefore,  $\ket{\widetilde{\Psi}}$ is a $U(g)$ invariant state identical to the original SPT ground state $\ket{\Psi}_{\mathrm{SPT}}$ within the simply connected region $R$. Therefore, we see that $\ket{\widetilde{\Psi}}$ must differ from $\ket{\Psi}_{\mathrm{SPT}}$ by the addition of a 2-dimensional SPT. Indeed, this SPT is characterized by the $3$-cocycle \eqnref{omega_x}, as can be seen by invoking dimensional reduction to (2+1)-D. Therefore, the original loop excitation corresponds to the boundary of such an SPT.

\section{Dimensional reduction: understanding loop excitations in terms of gapped boundaries between (2+1)-D topological phases}
\label{sec:degeneracy-from-topo}

In Section \ref{CC-in-DW}, we used a dimensional reduction argument to relate a
loop excitation in a $3+1$-dimensional Dijkgraaf-Witten topological phase with Abelian gauge
group $G$ to the boundary between a $2+1$-dimensional twisted Dijkgraaf-Witten topological phase and an untwisted topological phase with gauge group $G$. We now analyze such boundaries
in the case in which the $2+1$-dimensional theory is Abelian.

In an Abelian topological phase in $2+1$-dimenions, the fusion algebra of the particle types
is an Abelian group. Let us call this group $D$. Its elements are the particle types
and the group multiplication rule is the fusion rule.
The phase associated with braiding particles
of types $a$ and $a'$ is $e^{2\pi i b(a,a')}$,
where $b$ is a finite bilinear form $b: D\times D \rightarrow \mathbb{Q}/\mathbb{Z}$.
The different gapped boundary conditions \cite{Beigi11,Kapustin2011,Kitaev2012,Lindner2012,Clarke2013a,Barkeshli13b,Barkeshli13c,Barkeshli13d,Levin2013,
Kapustin2014a,Yoshida15,Cong16}
for the edge between this topological phase and the vacuum are given by the
different Lagrangian subgroups $L \subset D$. These are subgroups satisfying the
following two conditions: (1) for any $x,y \in L$, $b(x,y) = 0$; and (2) for any
$a \notin L$, there is some $x\in L$ such that $b(x,a) \neq 0$. In other words,
all of the particles in the Lagrangian subgroup braid trivially with each other
while every particle that is outside the Lagrangian subgroup braids non-trivially
with at least one particle in the Lagrangian subgroup. The second
condition implies that $|L| = \sqrt{|D|}$. The physical picture for the first condition
is that all of the particles in the Lagrangian subgroup can ``condense''
simultaneously at the edge, which confines all of the other particles, 
leading to a fully-gapped edge.

An annular region of the topological phase with two gapped boundaries, each of which is
associated with Lagrangian subgroup $L$, has an $|L|$-fold degenerate ground state.
One basis of these states is the set of eigenstates of topological charge on the outer boundary
(which must be compensated by the conjugate charge on the inner boundary).
There are $|L|$ such states since there is zero energy cost for any topological charge
in $L$ (since these charges have ``condensed'') and the energy cost for any topological charge outside $L$ is non-zero (since these charges are confined).\footnote{If the two gapped boundaries are associated with different Lagrangian subgroups, $L_1$ and $L_2$, then the
degeneracy is $|{L_1}\bigcap{L_2}|$.} In a Dijkgraaf-Witten gauge theory with finite
Abelian gauge group $G$, $|D| = |G|^2$: essentially, every particle can be labelled by
a charge, chosen from the $|G|$ irreps of $G$, and a flux, chosen from the $|G|$ elements
of the group. Hence, the ground state subspace has dimension $|L| = \sqrt{|D|} = |G|$.
As an example, consider the toric code. It has two Lagrangian subgroups:
${L_1} = \{1,e\}$ and ${L_2} = \{1,m\}$. Hence, there are two possible choices
of gapped boundaries of the annulus, each of
which is two-fold degenerate.
For doubled semions (the twisted
$\mathbb{Z}_2$ case), the particle types are $\{1, s, \overline{s}, b\}$ and
the only Lagrangian subgroup is $L = \{1, b\}$. Hence, there is a single choice
of gapped boundaries of the annulus, and it is two-fold degenerate.

In order to compute the degeneracy of a loop in a $3+1$-dimensional topological phase,
we need a slight modification of the preceding results since we need to consider boundaries between
a twisted Dijkgraaf-Witten theory $DW_\omega$ and an untwisted one $DW$, rather than a boundary into the vacuum. By folding along the boundary, this is equivalent to considering a boundary between $DW_\omega \times DW$ and the vacuum. Then we need to find a Lagrangian subgroup $L \subseteq A_\omega \times A$, where $A_\omega,A$ are the sets of particles in $DW_\omega,DW$ respectively.

First of all, we observe that $A$ can be written simply as $A = G \times G^*$, where $G$ labels the fluxes and $G^{*}$ (the group of 1-dimensional representations of $G$) labels the charges. The braiding in $DW$ of a charge $\chi \in G_*$ with a flux $g \in G$ is given by $\chi(g)$.
The structure of $A_\omega$ is a little bit more complicated. In fact, a particle in $A_\omega$ is labelled by $(g,\xi)$, where $g \in G$ and $\xi : G \to \U(1)$ satisfies 
\begin{multline}
\label{projrepcond}
\xi(g_1) \xi(g_1) \xi(g_1 g_2)^{-1} \\= \omega(g,g_1,g_2) \omega(g_1, g, g_2)^{-1} \omega(g_1, g_2, g).
\end{multline}
The mutual statistics of a particle $(g_1,\xi_1)$ with a particle $(g_2,\xi_2)$ is given by $\xi_2(g_1) \xi_1(g_2)^{-1}$, while the self-statistics of a particle $(g,\xi)$ is $\xi(g)$. We note that it is always possible to choose a normalization of $\omega$ such that the RHS of \eqnref{projrepcond} is trivial for $g=1$. Thus, the complication here is in the dyons -- particles carrying both flux and charge. If we focus on the ``pure charge'' sector $g=1$ then $\xi$ just becomes a 1D representation of $G$.

Hence the excitations of $A_\omega \times A$ can be denoted by
$(g,\xi)\times (k,\chi)$ where $g,k \in G$, $\xi$ satisfies Eq. (\ref{projrepcond}), and $\chi$
is an irrep of G. If our sole concern were the possible gapped edges of
$A_\omega \times A$, then our task would simply be to find the Lagrangian
subgroups of $A_\omega \times A$. But our interest here is gapped loop
excitations in 3+1 dimensions, so we want only want those Lagrangian subgroups $L$
that satisfy extra conditions that guarantee that they answer the question of interest
in one higher dimension.
All ``pure charge'' excitations should pass through the edge separating $A_\omega$ from $A$
since, in 3+1-D, a charge can simply go around a loop to get from one side to the other.
Hence, $L$ must contain all elements of the form $(1,\chi)\times (1,\chi)$; when such a particle
condenses on the boundary between $A_\omega$ and $A$, $(1,\chi)$ in $A_\omega$ can freely
convert to $(1,\chi)$ in $A$.
In fact, some fluxes in 3+1-D may be able to pass through
the loop unaffected, which means that there may be some subgroup
$H\subset G$ such that
$(h,\chi)\times (h^{-1},\chi)$ condenses for all $h\in H$.
(The ``pure charges'' are simply a special case of this with zero flux.)
Here, the particles on both sides of
the boundary must carry the same charge $\chi$, since the presumption is that these flux
loops can move unchanged through the flux loop that defines $A_\omega$
which, in turn, presumes that the 3-cocycle $\omega$ characterizing
$A_\omega$ is trivial when restricted to $H$.

Now we claim that we can construct a Lagrangian subgroup $L \leq A_\omega \times A$
satisfying the requirements stated above given any subgroup $H \leq G$ such that $\omega$ becomes trivial when restricted to $H$. The subgroup $L$ consists of all particles of the form $(h,\xi) \times (h^{-1},\chi)$ where $(h,\xi) \in A_{\omega}$, $(h^{-1},\chi) \in A$, and $\chi(h') = \xi(h')$ for all $h' \in H$. The condition that $\omega$ becomes trivial when restricted to $H$ is necessary to make these conditions compatible with \eqnref{projrepcond}.
The interpretation of $H$ is the set of fluxes that are allowed to pass through the boundary. Moreover, the set of ``pure charges'' that can annihilate on the boundary is
$C = \{ \chi \in G_* | \chi(h) = 1 \quad \forall h \in H \}$.

 To prove that $L$ is a Lagrangian subgroup, first note that the braiding between two particles $(h_1,\xi_1)(h_1^{-1}, \chi_1)$ and $(h_2,\xi_2)(h_2^{-1}, \chi_2)$ is given by
\begin{multline}
\xi_1(h_2) \xi_2(h_1)^{-1} \chi_1(h_2^{-1}) \chi_2(h_1^{-1})^{-1} \\=
[\xi_1(h_2) \chi_1(h_2)^{-1}] [\chi_2(h_1) \chi_1(h_2)^{-1}]^{-1} = 1.
\end{multline}
Similarly one finds that the self-statistics of a particle $(h,\xi)(h^{-1},\chi)$ are given by
\begin{equation}
\xi(h^{-1}) \chi(h) = \xi(h)^{-1} \chi(h) = 1.
\end{equation}

Next we need to prove that any particle which braids trivially with $L$ must be contained in $L$. We will need the following Lemma:
\begin{lemma} (a) Suppose that $g \in G$ satisfies $\chi(g) = 1$ for all $\chi \in G^{*}$. Then $g = 1$. (b) Suppose that $g \in G$ satisfies $\chi(g) = 1$ for all $\chi \in C$. Then $g \in H$.
\begin{proof}
First of all we note that (a) is a special case of (b) if we set $H$ to be the trivial subgroup. To prove (b), we define $H' = \{ g \in G : \chi(g) = 1 \quad \forall \chi \in C \}$. $H'$ can be interpreted as the kernel of a homomorphism $\varphi : G \to C^{*}$, $\varphi(g)[\chi] = \chi(g)$. Composing with the natural isomorphisms $C^{*} \cong (G/H)^{**} \cong G/H$ we obtain precisely the natural projection map $G \to G/H$. So we conclude that $H' = H$.
\end{proof}
\end{lemma}

Now suppose we have a particle $(g,\xi)(g',\chi) \in A_\omega \times A$ which braids trivially with $L$. Braiding with particles of the form $(1,\chi)(1,1)$ and $(1,1)(1,\chi)$ (where $\chi \in C$) shows that $g,g' \in H$. Braiding trivially with particles of the form $(1,\chi)(1,\chi)$ (where $\chi \in G^{*}$) shows that $g' = g^{-1}$. Finally, braiding trivially with particles of the form $(h,\xi_1)(h^{-1},\chi_1)$ shows that $\xi(h) = \chi(h)$ for all $h \in G$.

What is the degeneracy associated with the boundary? To answer this, we consider the system on a sphere with a boundary between $DW$ and $DW_{\omega}$ on the equator. We want to think of the sphere as really being the infinite plane, where one point $x_*$ on the sphere is at infinity. Therefore, we need to allow some anyon $a \in A$ to be located at $x_*$.
By flattening the sphere to a pancake, we can also imagine that the equator is a boundary between $DW_\omega \times DW$ and the vacuum, with a charge $a$ somewhere in the bulk. (Now we think of $a$ as a particle type in the layered theory $DW_\omega \times DW$, but we remember that it is certainly trivial in one of the two layers.) This is allowed configuration only if $a \in L$, which given our definition of $L$ is equivalent to saying that $a$ is a charge in $C$. Therefore, the topological degeneracy of the boundary is $|C|$, correponding to the different Cheshire charges that can exist on the boundary.

\section{Equivalence of two forms of data characterizing loop excitations}
\label{forms_equivalence}
We can also specify the data for a loop-like excitation in a Dijkgraaf-Witten theory, as described in Section \ref{dimreduction_enumerate}, can also be characterized in a mathematically equivalent way, which will prove more convenient for our discussion of membrane-net models and category-theoretical interpretations. Let $\{ \mu : \mu \in M \}$ be a set of labels for the symmetry-breaking states, and let the symmetry define an action $\mu \mapsto g \mu$ of $G$ on these labels. We assume that the action is \emph{irreducible}; that is, there is no proper subset of $M$ that is left invariant by the action. Then we claim that the loop-like excitations carrying flux $x$ can also be specified by the following data:
\begin{framed}
\textbf{Data for loop-like excitations carrying flux $x$. (Form II)}
\begin{itemize}
\item A set $M$, and an irreducible action of $G$ on $M$.
\item A function $\alpha : M \times G \times G \to U(1)$ [written $\alpha^\mu(g_1, g_2)$] such that
\begin{equation}
\label{anomaly_matching}
(d\alpha)^\mu (g_1, g_2, g_3) = \widetilde{\omega}_x(g_1, g_2, g_3),
\end{equation}
where we identify $\alpha$ and $\alpha^{\prime}$ if they differ by $\alpha = \alpha^{\prime} \times d\beta$ for some function $\beta^\mu(g)$.
\end{itemize}
\end{framed}
Here we have introduced the generalized coboundary operators
\begin{equation}
\label{generalized_coboundary}
(d \alpha)^{\mu}(g_1, g_2, g_3) =
\frac{\alpha^{g_1 \mu} (g_2, g_3) \alpha^{\mu}(g_1, g_2 g_3)}{\alpha^{\mu}(g_1, g_2) \alpha^{\mu}(g_1 g_2, g_3)},
\end{equation}
\begin{equation}
(d\beta)^{\mu}(g_1, g_2) = \frac{\beta^{g_1 \mu}(g_2) \beta^{\mu}(g_1)}{\beta^{\mu}(g_1 g_2)}.
\end{equation}
Eq.~(\ref{anomaly_matching}) can be interpreted loosely as a ``anomaly matching'' condition: in a truly (1+1)-D system, $d \alpha = 1$, and if $d \alpha \neq 1$ this represents an anomaly which must be cancelled by the contribution $\widetilde{\omega}_x$ from the bulk. The topological degeneracy is the size of the set $M$.

We call the original form of the data, as given in Section \ref{dimreduction_enumerate}, ``Form I''. Here we prove that Forms I and II are in one-to-one correspondence. Going back and forth between the invariant subgroup $H$ and irreducible permutation action of $G$ on a set $M$ is straightforward.

We will then invoke the following Lemma:
\begin{lemma}
\label{shapiro}
Let $\U(1)^M$ denote the $G$-module of functions $f : M \to \U(1)$, with $g$ acting by permutation on $M$, i.e. $(gf)(\mu) = f(g\mu)$. Then the map from $\mathcal{H}^n(G, \U(1)^M) \to \mathcal{H}^n(H, \U(1))$ induced by restriction onto $H$ is an isomorphism.
\begin{proof}
This is a special case of Shapiro's Lemma \cite{Weibel1995}.
\end{proof}
\end{lemma}
We then observe that \eqnref{anomaly_matching} has a solution for $\alpha^\mu$ if and only if the right hand side, thought of as a $\U(1)^M$-valued cochain, corresponds to the trivial element of $\mathcal{H}^3(G, \U(1)^M)$. By Lemma \ref{shapiro}, this is equivalent to requiring that the restriction of $\widetilde{\omega_x}$ onto $H$ is trivial. Furthermore, the different solutions are a torsor over $\mathcal{H}^2(G, \U(1)^M)$, which by Lemma \ref{shapiro} is equivalent to $\mathcal{H}^2(H, \U(1))$. This completes the proof of the one-to-one correspondence between Forms I and II.

\section{Membrane-net models}
\label{membranenet}
In Section \ref{sec:lattice-model}, we showed how to construct membrane operators that create excitations, in a particular model of a $\mathbb{Z}_2 \times \mathbb{Z}_2$ gauge theory. In this section, we outline an alternative perspective on such membrane operators that we expect to be applicable to any Dijkgraaf-Witten theory in (3+1)-D, or even to more general topological phases. This approach will also allow us to make explicit the category theory formulation of the classification of the excitations (see next section).
The idea is to consider the analogue of Levin and Wen's string-net models in higher dimensions. In (3+1)-D, the ``string-net condensate'' of Levin and Wen becomes a ``membrane-net condensate''. In (2+1)-D, any Abelian Dijkgraaf-Witten theory (indeed, we conjecture, any Dijkgraaf-Witten theory whatsoever) can be realized in a suitable string-net model, and we expect this to be true in higher dimensions as well. Presumably, one can also construct more general topological phases through such a membrane-net construction as well (though we will not attempt to do this in the present paper.) Our discussion will be sketchy in parts, as our aim is to demonstrate the main concepts and show that the classification of the excitations from Section \ref{dimreduction_enumerate} arises naturally in this context, rather than to spell out all the details.

\begin{figure}
    \input{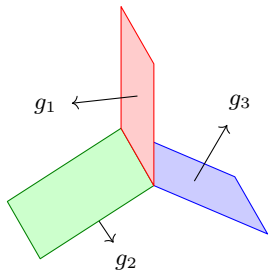}
    \caption{\label{membraneboundary}The product $g_1 g_2 g_3$ around a three-fold intersection between
    membranes must be $1$.}
\end{figure}

A membrane-net model is formulated in terms of spins on the faces of some three-dimensional lattice. We let $\mathcal{S}$ be a set labelling the basis of each spin. We can then imagine that each basis state $\ket{s_1, s_2, \cdots}$, $s_j \in \mathcal{S}$ represents a configuration of membranes labelled by the elements of $\mathcal{S}$, where a special element $1 \in \mathcal{S}$ denotes the absence of a membrane. For each $s \in \mathcal{S}$, we choose an element $s^{*}$ to be its ``dual'' (with $1^{*} = 1$). We choose some orientation (i.e. unit normal vector) for all the faces of the lattice, and identify membrane configurations that differ by reversing the orientation of a face and changing its label from $s \to s^{*}$.
 We assume that the lattice is chosen such that every edge is shared between
 exactly three faces, and the allowed membrane-net configurations are determined
 by the \emph{branching rules}, which determine what three membrane types are
 allowed to meet at an edge. In this paper, we will just consider the case where
 the label set $\mathcal{S}$ is actually a group $G$, and the branching rules
 require that the product of labels around an edge be 1. (See Figure
 \ref{membraneboundary}).
 However, one could also consider more general branching rules, as Levin and Wen did in the (2+1)-D case.

\begin{figure}
\input{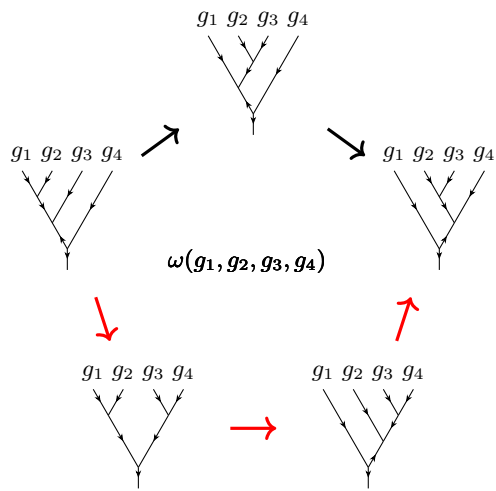}
\caption{\label{pentagon} The pentagon move relates two different membrane-net configurations (shown with black and red connecting arrows respectively.)}
\end{figure}

The ground state wavefunction of the model is a superposition over membrane-net configurations:
\begin{equation}
\ket{\Psi_0} = \sum_{\mbox{$\mathcal{C}$}} \Psi(\mathcal{C}) \ket{\mathcal{C}},
\end{equation}
where the sum is over all membrane-net configurations allowed by the branching rules. The amplitude $\ket{\Psi(\mathcal{C})}$ is determined implicitly by a set of rules relating the amplitudes of configurations differing by some basic moves. It is not necessary for our purposes here to specify the complete set of rules and the consistency conditions that they must satisfy, and we leave this for future work. However, certainly there must be some rule associated with the \emph{pentagon move} depicted in Figure \ref{pentagon}, where the pictures represent two-dimensional slices through a three-dimensional membrane configuration, and arrows connect adjacent slices\footnote{The arrows indicate the orientation of the membranes, by taking the cross product with some choice of normal vector to the slices.} We assume that the amplitudes for the membrane-net configurations depicted in Figure \ref{pentagon} are related by multiplication of a phase factor $\omega(g_1, g_2, g_3, g_4)$. The fact that the wavefunction must be single-valued imposes the requirement that $\omega$ be a 4-cocycle (this follows from requring the consistency of the three-dimensional ``associahedron'' diagram; for example, see Figure 18 of Ref.~\onlinecite{Kitaev2006a}). We expect a membrane-net model with such a pentagon rule to correspond to a Dijkgraaf-Witten theory with the same 4-cocycle.

We will now explain how to construct membrane operators that leave such a membrane-net wavefunction invariant. The idea is inspired by the analogous \emph{string operators} constructed by Levin and Wen in (2+1)-D. The construction we use differs from the one in the Section \ref{sec:lattice-model} (and resembles the construction of Levin and Wen) in that we construct the operators \emph{explicitly} rather than through implicit rules. However, the operators that we construct obey implicit rules similar to those of Section \ref{sec:lattice-model}, so our method can be thought of as finding explicit solutions to the implicit rules of Section \ref{sec:lattice-model}.

We construct the membrane operator as follows: let $S$ be an oriented surface such that the intersections with the faces of the lattice are generic (such that, for example, the surface does not touch a face of the lattice without passing through it.) We imagine starting from a membrane configuration $\mathcal{C}$, then laying down an extra membrane of type $x$ on $S$. We denote the resulting membrane configuration by $\mathcal{C} + x_S$. (There is some ambiguity in defining this configuration due to the need to choose a resolution of the intersections; all that matters is that we can choose a consistent convention depnding on the orientation of $S$ and the faces it intersects with). We then define the resulting state by
\begin{equation}
\Psi(\mathcal{C} + x_S) = A(\mathcal{C} + x_S) \Psi_0(\mathcal{C}),
\end{equation}
where $A$ is some coefficient to be determined, and on the left-hand side it is understood that we use the basic membrane-net rules to reduce $\mathcal{C} + x_S$ to a configuration in which membranes live only on the faces of the lattice.

We now adopt two complementary graphical representations. In the first, we draw two-dimensional pictures representing portions of the surface $S$, with lines corresponding to the intersections of membranes with $S$.
These pictures will be shaded blue to indicate that they depict only the configurations adjacent to the surface $S$.
We also have another graphical representation in which we draw a series of two-dimensional slices of three-dimensional space in planes intersecting the surface $S$ (say, planes of constant $z$ coordinate). The interection of these slices with the membrane $S$ shown by a thick blue line, and the arrows on the lines in the picture of the slice indicate the orientation of the corresponding surfaces via a cross product with the normal vector to the slice.
 To illustrate these two representations, here is the same configuration in both representations:
\begin{equation*}
%
%
%
\input{V.fig}
\end{equation*}
\begin{equation}
\label{V_move}
  \input{move_over_vertex.fig}
\end{equation}
Here V denotes some particular resolution of the quadruple intersection point.

The coefficient $A(\mathcal{C} + x_S)$ is then evaluated by taking a product over a factor $\alpha(g_1, g_2)$ for each quadruple intersection point involving $S$, as shown in \eqnref{V_move}
In particular, we assign this factors to a particular ``canonical'' resolution V of the quadruple intersection point (for any other resolutions, we first use the membrane-net rules to reduce to a canonical resolution.) Henceforth, in pictures such as \eqnref{V_move}, we will assume that these extra phase factors $\alpha(g_1,g_2)$ have been included.


\begin{figure}
\input{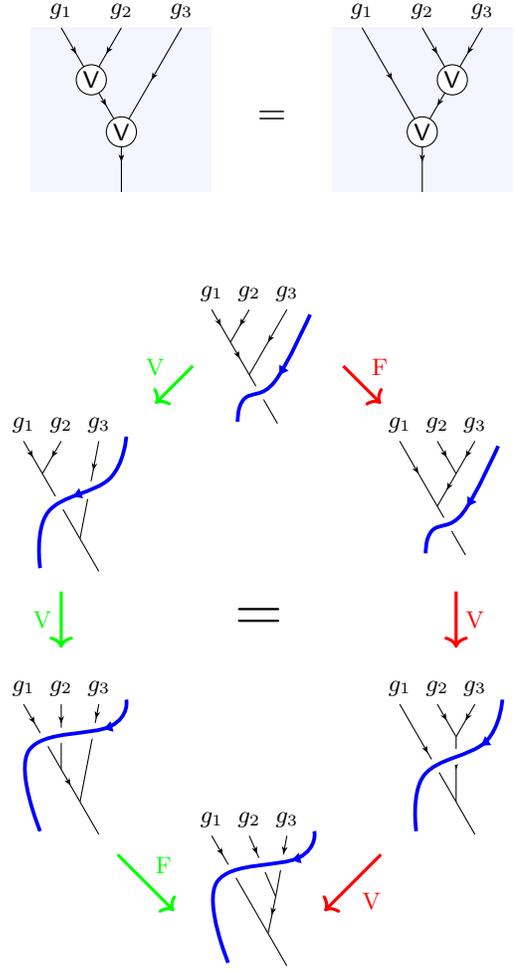}
\caption{\label{Fthrough}The consistency condition for the membrane operator, in the two different graphical representations described in the text.}
\end{figure}
We require that the membrane operator, when defined on a closed surface $S$, leaves the ground-state wavefunction $\Psi_0$ invariant. In order for this to be true we must impose the condition shown in Figure \ref{Fthrough}. In Appendix \ref{membrane_derivation}, we show that this in turn implies (for a particular choice of the canonical resolution $V$; see the Appendix for details) that
\begin{equation}
\label{hexagon_cond}
d\alpha = \widetilde{\omega}_x,
\end{equation}
where $d$ is the coboundary operator:
\begin{equation}
(d\alpha)(g_1, g_2, g_3) = \frac{\alpha(g_2, g_3) \alpha(g_1, g_2 g_3)}{\alpha(g_1, g_2) \alpha(g_1 g_2, g_3)},
\end{equation}
and $\widetilde{\omega}_x$ is the 3-cocycle defined in \eqnref{omega_x}.

\eqnref{hexagon_cond} poses a conundrum, because we we already saw in Section \ref{dimreduction_enumerate} above that the 3-cocycle $\widetilde{\omega}_x$ appearing on the right-hand side is in general not trivial, and thus cannot be written as a coboundary as \eqnref{hexagon_cond} would suggest. Thus, the membrane operator that we constructed by the procedure above can only produce the \emph{Abelian } loop excitations. These membrane operators are the equivalent of the string operators considered in Ref.~\onlinecite{Lin2014}, which similarly could only describe Abelian  particles.

Fortunately, we can easily resolve this problem by considering a simple generalization. We introduce a new label set $M$, and an action $\mu \mapsto g\mu$ of $G$ on $M$.
We label patches of the membrane $S$ by elements of $M$, in such a way that patches separated by a membrane of type $g$ differ in their labels by the action of $g$, e.g. 
\begin{equation}
\label{Vmod}
\input{Vmod.fig}
\end{equation}
Then we generalize $\alpha(g_1, g_2)$ to $\alpha^{\mu}(g_1, g_2)$ which describes the phase factor corresponding to the configuration shown in \eqnref{Vmod}. One can show by a simple generalization of Appendix \ref{membrane_derivation} that $\alpha^\mu$ must satisfy the condition
\begin{equation}
(d\alpha)^{\mu}(g_1, g_2, g_3) = \widetilde{\omega}_x(g_1, g_2, g_3),
\end{equation}
where $d$ is the generalized coboundary operator defined in \eqnref{generalized_coboundary}. Since this is exactly the condition characterizing the types of loop excitations (see Section \ref{dimreduction_enumerate} and Appendix \ref{forms_equivalence}), we expect that such excitations are created using the membrane operators that we have just constructed. The choice of label $\mu$ presumably corresponds to the topological degeneracy of the excitations, though we will not attempt to prove this. 

\section{Derivation of the consistency condition for membrane operators}
\label{membrane_derivation}
\begin{figure}
\includegraphics[scale=0.22]{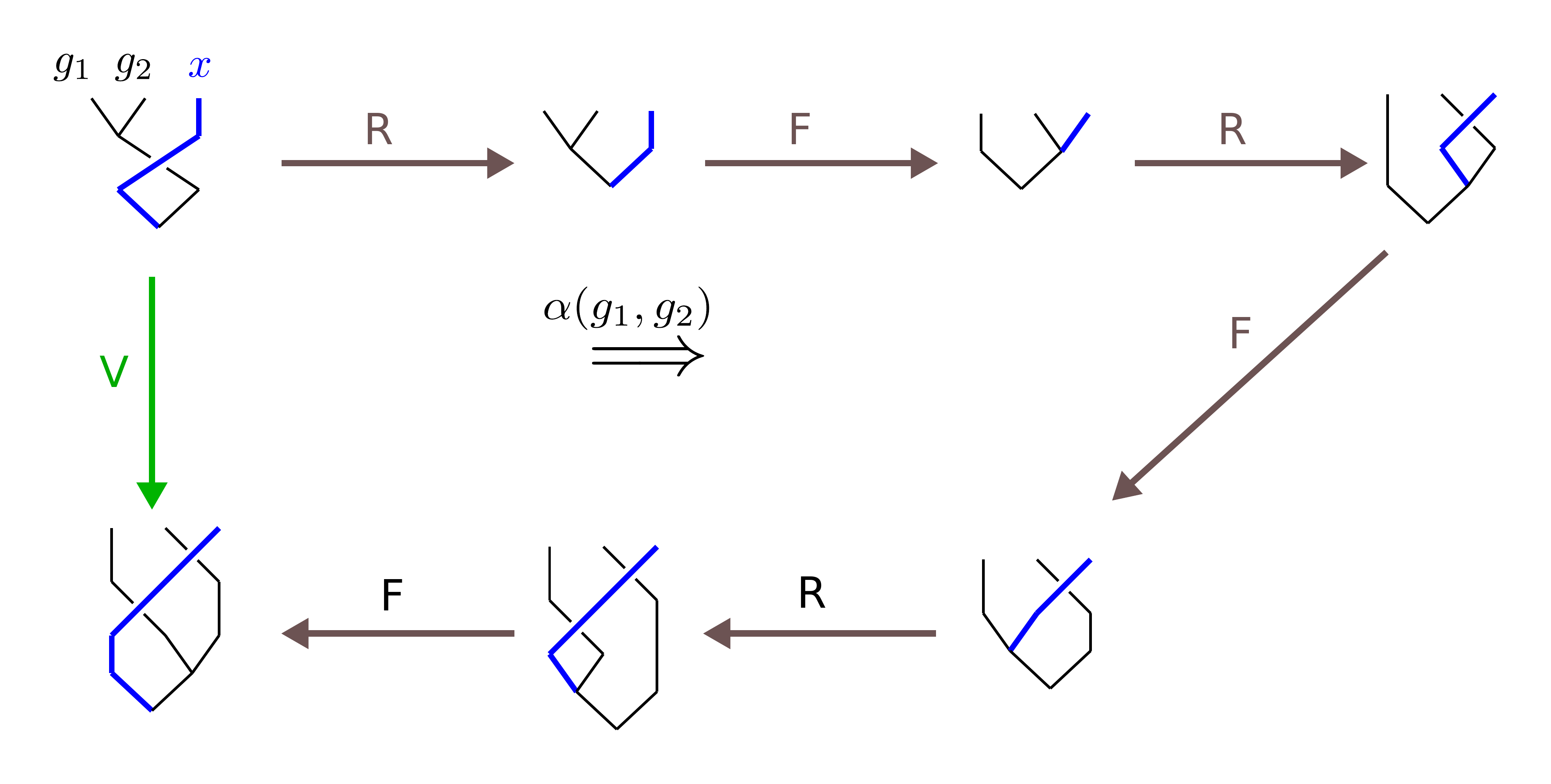}
\caption{\label{hexagon_move}The hexagon move.}
\end{figure}

Here we will show how the consistency condition \eqnref{hexagon_cond} follows from Figure \ref{Fthrough}. First we have to make some choice of the ``canonical'' resolution $V$. We will not specify it explicitly but rather fix it by the requirement that the ``hexagon move'', as depicted in Figure \ref{hexagon_move}, must evaluate to $\alpha(g_1, g_2)$ (where we also included the extra $\alpha(g_1, g_2)$ factor in the definition of $V$; otherwise, the hexagon move would evaluate to 1.) Here $R$ can be chosen to be any canonical resolution of the corresponding configuration, so long as we are consistent throughout.

\begin{figure*}[p]
\includegraphics[scale=0.18]{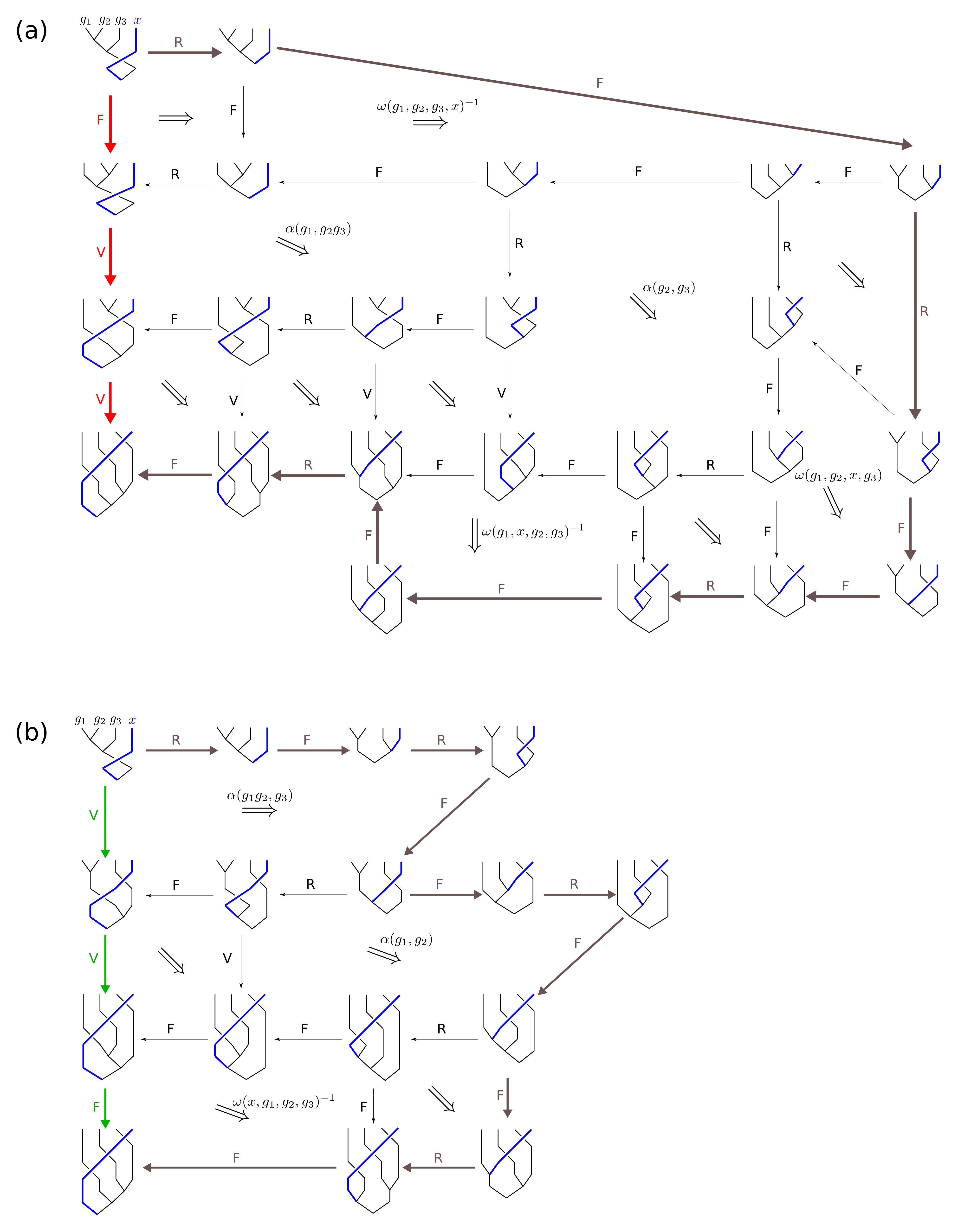}
\caption{\label{hexagon_cons_big}In parts (a) and (b), each of the two membrane-net configurations equated by the consistency condition Figure \ref{Fthrough} (shown with red and green arrows) is related to the \emph{same} membrane-net configuration shown with gray arrows by a sequence of pentagon and hexagon moves.}
\end{figure*}

The derivation of \eqnref{hexagon_cond} then follows from Figure\ref{hexagon_cons_big}.

\section{The definition of the category $\cat{2Vect}_G^{\omega}$}
\label{sec:def-2Vect}

Here we will explain how to define the fusion 2-category $\cat{2Vect}_G^{\omega}$ mentioned in Section \ref{sec:2-categories}. But first, as a warm-up, we define the fusion category $\textbf{Vect}_G^{\omega}$, which is the category of $G$-graded vector spaces, twisted by a \emph{3-cocycle} $\omega$. This fusion category is the input into a Levin-Wen construction in which string types are labelled by elements of $G$ and the fusion rules are simply group multiplication. The monoidal center $Z(\textbf{Vect}_G^{\omega})$ describes the point excitations in a (2+1)-D Dijkgraaf-Witten theory. We assume the reader is familiar with the concept of a fusion category, e.g.~see Refs.~\onlinecite{Bakalov2001,Etingof2005,Kitaev2006a}.

A $G$-graded vector space is a direct sum $\bigoplus_{g \in G} V_g$  of vector
spaces $V_g$ for each group element $g \in G$. We write $[V]_g$ for the graded
vector space in which only one term in the sum is non-trivial and equal to the
vector space $V$. The structure of the category of $G$-graded vector spaces is largely inherited from $\cat{Vect}$, the category of vector spaces. In particular, we define the  define the tensor product on $G$-graded vector spaces according to $[V]_g \otimes [W]_h = [V \otimes W]_{gh}$, or more generally
\begin{equation}
\left(\bigoplus_{g \in G} V_g\right) \otimes \left(\bigoplus_{h \in G} V_h\right) = \bigoplus_{k \in G} \left(\bigoplus_{g,h \in G : gh = k} V_{g} \otimes V_h\right).
\end{equation}
We define a \emph{$G$-graded linear map} between $G$-graded vector spaces $V$ and $W$ to be a block-diagonal sum $\bigoplus_g L_g$ of linear maps $L_g : V_g \to W_g$. We use a similar square bracket notation as above, $[L]_g$ to denote a $G$-graded linear map in which only one term in the sum is non-trivial. The tensor product of $G$-graded linear maps is defined as for $G$-grade vector spaces, e.g.~$[L]_g \otimes [K]_h = [L \otimes K]_{gh}$. This allows us to define a monoidal category of $G$-graded vector spaces. However, we have some freedom in the choice of the associator natural isomorphism $\alpha$. Indeed, one can verify that for any choice of $3$-cocycle $\omega$ we can take
\begin{multline}
\eta( [V]_g, [W]_h, [U]_k ) : [(V \otimes W) \otimes U]_{ghk} \to [V \otimes (W \otimes U))]_{ghk} \\=
\omega(g,h,k) 
\times[\eta_{V,W,U}]_{ghk},
\end{multline}
 where $\eta_{V,W,U}$ is the associator in $\textbf{Vect}$, the category of (not $G$-graded) vector spaces. (The condition that $\omega$ be a 3-cocycle comes from the pentagon identity.) This gives the category $\cat{Vect}_G^{\omega}$. Strictly speaking, we have only constructed it as a weak monoidal category; however, it can be shown that it can be equipped with additional structure to make it a unitary fusion category.
 
Now we can similarly construct the fusion 2-category $\cat{2Vect}_G^{\omega}$ for a 4-cocycle $\omega$. Actually, we will only construct it as a weak monoidal 2-category. We expect that it can be endowed with suitable additional structure to make it a fusion 2-category, but we will not prove this.

A 2-vector space is a ``categorification'' of the notion of a vector space \cite{Kapranov1991}. Specifically, whereas the elements of an $n$-dimensional vector space can be thought of as a column vector of $n$ complex numbers, an element of an $n$-dimensional 2-vector space is a column vector
\begin{equation}
\begin{bmatrix}
V_1 \\
V_2 \\
\cdots \\
V_n \\
\end{bmatrix},
\end{equation}
where \emph{each} $V_i$ is a vector space. Whereas a morphism in $\cat{Vect}$, the category of vector spaces, can be thought of as a matrix $\alpha_{ij}$ of complex numbers, the morphisms in $\cat{2Vect}$, the category of 2-vector spaces are matrices $A_{ij}$, where \emph{each entry} $A_{ij}$ is a vector space. These matrices act on elements of the 2-vector space by ``matrix multiplication'', i.e. $(AV)_i = \bigoplus_k A_{ik} \otimes V_k$. Finally, a 2-morphism in $\cat{2Vect}$ between morphisms $A$ and $B$ is a matrix of linear maps $L_{ij} : A_{ij} \to B_{ij}$. 

We can now define a $G$-graded 2-vector space in a similar way as we did above for vector spaces. The category of $G$-graded 2-vector spaces then inherits most of its structure from $\cat{2Vect}$. However, we still have some freedom in defining the ``pentagonator'' $P$ that appears in the definition of a weak monoidal 2-category. In general, we can have
\begin{equation}
P([\mathcal{V}_1]_{g_1}, [\mathcal{V}_2]_{g_2}, [\mathcal{V}_3]_{g_3}, [\mathcal{V}_4]_{g_4}) = \omega(g_1, g_2, g_3, g_4) P(\mathcal{V}_1, \mathcal{V}_2, \mathcal{V}_3, \mathcal{V}_4),
\end{equation}
where the $P$ appearing on the right-hand side is the pentagonator of $\cat{2Vect}$, and $\omega$ is a 4-cocycle. This gives the category $\cat{2Vect}_G^{\omega}$.

%

\bibliography{../References/references}
\end{document}